\documentclass{jfm}
\pdfoutput=1
\usepackage{graphicx}
\usepackage{subfigure}
\usepackage{epstopdf,epsfig}
\usepackage{amsmath}
\usepackage{natbib}
\usepackage{bm}
\usepackage{tikz}
\usepackage{lineno}
\usepackage{mathrsfs}
\usepackage{mathtools}
\usepackage[center]{caption}
\allowdisplaybreaks[1]
\usepackage{hyperref}
\hypersetup{colorlinks,
	citecolor=blue}

\shorttitle{Third-order solution of surface gravity waves} 
\shortauthor{Zhe Gao} 

\title{On two approaches to the third-order solution of surface gravity waves}

\author{Zhe Gao
  \corresp{\email{gaozhe@mail.dlut.edu.cn}},
  Z.C Sun
 \;\and
  S.X Liang
}
\affiliation
{
	State Key Laboratory of Coastal and Offshore Engineering, Dalian University of Technology, Dalian, 116024, PR China
}

\begin{document}
\maketitle

\begin{abstract}
	Third order approximate solution for surface gravity waves in the finite water depth are studied in the context of potential flow theory. This solution corresponds to the bound harmonics of a multidirectional irregular wavefield, and provides explicit expressions for the surface elevation, free-surface velocity potential and velocity potential. The amplitude dispersion relation is also provided. Two approaches are used to derive the third order analytical solution, resulting in two types of approximate solutions: the perturbation solution and the Hamiltonian solution. The perturbation solution is obtained by classical perturbation technique in which the time variable is expanded in multiscale to eliminate secular terms. The Hamiltonian solution is derived from the canonical transformation in the Hamiltonian theory of water waves. By comparing the two types of solutions, it is found that they are completely equivalent for the first to second order solutions and the nonlinear dispersion, but for the third order part only the sum-sum terms are the same. Due to the canonical transformation that could completely separate the dynamic and bound harmonics, the Hamiltonian solutions break through the difficulty that the perturbation theory breaks down due to singularities in the transfer functions when quartet resonance criterion is satisfied. Furthermore, it is also found that some time-averaged quantities based on the Hamiltonian solution, such as mean potential energy and mean kinetic energy, are equal to those in the initial state in which sea surface is assumed to be a Gaussian random process. This is because there are associated conserved quantities in the Hamiltonian form. All of these show that the Hamiltonian solution is more reasonable and accurate to describe the third order steady-state wave field. Finally, based on the Hamiltonian solution, some statistics are given such as the volume flux, skewness, excess kurtosis, and non-uniqueness of induce mean flow and mean surface. 

\end{abstract}


\section{Introduction}
Surface gravity waves have always been a fascinating subject in the field of ocean engineering and applied mathematics, which are usually wind-driven and propagate on ocean surface under the influence of gravity, mainly including wind wave and swell. For a real sea state, the interaction between numerous primary waves forms a random wave field characterized by irregularity and directional spreading. It is well known that the random wave field in infinite or finite water depth is a dynamic evolution process in which quartet resonances lead to energy exchange between wave components. Before the quartet resonance  mechanism was discovered by \cite{phillips1960dynamics}, steady waves had been extensively studied. Unlike the dynamic wave system, the amplitude of steady wave is time-independent. The study of steady waves can be traced back to the Airy wave theory, also referred to as linear wave theory, which gives a linearized description of random wave field by the superimposition of a large number of elementary waves having different wavelengths, frequencies, and directions of propagation. Linear wave theory is often applied to ocean engineering and coastal engineering for many purposes by giving a description of the wave kinematics and dynamics \citep[see e.g.,][]{mei2018theory,Massel2018Ocean}. Moreover, the linear random wave model is a cornerstone in wave statistics theory \citep[see e.g.,][]{goda2010random,ochi2005ocean}. For example, in a linear random wave model, the surface elevation distribution follows the Gaussian law and the wave height follows the Rayleigh distribution for an infinitely narrow spectrum \citep{longuet1952statisticaldistribution}. But the linear wave theory is only accurate for infinitesimal amplitudes. When the amplitude of steady waves is finite, the nonlinear interaction of elementary waves must be considered. Thus, it is necessary to seek the nonlinear steady-state solution of multidirectional irregular waves.

Potential flow theory enables us to describe random wave field by using irrotational Euler equations. Although the governing equation (i.e., Laplace equation) is linear, the exact analytic solution is still unknown due to the nonlinear boundary conditions, even in the simplest case of travelling waves of permanent form. Therefore, many researchers appeal to approximate method to solve this water wave problem. The classic perturbation expansion theory has been one of the most powerful tools for solving general nonlinear equations. Since the pioneering work of \cite{Stokes47,stokes1880supplement}, the progressive periodic waves of permanent form in finite or infinite depth, i.e., Stokes wave, has been extensively studied by using the perturbation expansion technique, such as the fifth-order Stokes wave \citep{skjelbreia1960fifth,fenton1985fifth}. Because the analytical derivation of higher order Stokes wave becomes difficult, numerical computations have drawn much attention of many researchers \citep{schwartz1974computer,fenton1988numerical,clamond_dutykh_2018}, unfolding important nonlinear characteristics such as the non-monotonic behaviors of the integral properties against wave steepness near the limiting Stokes wave \citep{longuet1975integral}. However, although the Stokes wave theory reveals many nonlinear properties of surface gravity waves, it is only a theory of monochromatic periodic wave without considering the irregularity and directional spreading of random wave field. To develop the analytical theory of nonlinear random waves, the nonlinear interaction between different primary waves should be considered. As the first step of studying random waves, the monochromatic short-crested wave has been investigated both analytically and numerically \citep{hsu1979third,roberts1983highly,okamura2010almost}.

The analytic theory of multidirectional irregular waves is much less developed perhaps due to the increased complexity. In principle, in the derivation of approximate analytical solutions of multidirectional irregular waves, one should consider the double interaction of primary waves as a kernel  for second-order solutions, and the triple interaction for third-order solutions. For instance, the third-order solution is composed of a triple summation over all possible pairs of wave components, based on a third-order solution for tridirectional trichromatic waves as a kernel in the summation. Using the perturbation expansion technique to solve irrotational Euler equations, the second-order solution was first derived by \citet{longuet1963effect} for random waves in deep water, and later by \citet{sharma1981second,dalzell1999note} for finite water depth, which is called the second-order random wave theory. A third-order perturbation solution for unidirectional irregular waves in deep water was attempted by \citet{pierson1993oscillatory} with an incorrect amplitude dispersion relation, and later his result was revised by \citet{zhang1999general}. On the basis of the third-order solution for bidirectional bichromatic waves \citep{madsen2006third}, \citet{madsen2012third} derived a third-order solution for multidirectional irregular waves in finite water depth with the option of specifying zero net volume flux. In fact, a fourth-order solution for nonlinear interactions among multiple directional wave trains has already been derived by \citet{ohyama1995fourth}, who showed that nonlinear components cause isolated large waves in the random wave field, especially the fourth-order component significantly contribute to bound low-frequency waves. But this conclusion is questionable because the perturbation solution will collapse near the resonance point, as pointed out by \citet{madsen2012third}, resulting in an unreliable surface elevation. 

Recently, the difficulties of singularities in the perturbation solution have been overcome by means of the homotopy analysis method (HAM) in the study of steady-state resonant waves \citep{liao2011homotopy,xu2012steady,liu2014steady,liao2016steady,liu2018finite}. HAM also successfully solved the highest Stokes wave in arbitrary water depth \citep{zhong2018limiting}. These studies confirm the flexibility of HAM without depending on small physical parameters. Although the steady-state resonance wave they studied involves only two primary waves which is a special case in the steady wave system that satisfies exact or near resonance criterion, there is no doubt that these can be extended to random wave fields involving any number of primary waves. Nevertheless, while increasing the amount of calculation, HAM cannot give the explicit expression for the transfer function in nonlinear solutions.

Different from the perturbation theory using original water wave equations as a starting point, the Hamiltonian theory of water wave was established by \citet{zakharov1968stability} who asymptotically expanded the Hamiltonian of water wave system in terms of sea surface and free-surface velocity potential, and removed the bound harmonics resulting in an integro-differential equation with quartet resonances, known as the Zakharov equation. The extension of Zakharov equation including quintet resonances was derived by \cite{crawford1980evolution,stiassnie1984modifications} for finite water depth. But there is a fundamental shortcoming that their versions of Zakharov equation is not Hamiltonian (i.e., not non-conservative), whereas the original water wave equations are conservative. This is because the multiple scale method was employed to separate free and bound harmonics, and their derivation takes the original water wave equation as the starting point, not the Hamiltonian formulation. \citet{krasitskii1994reduced} overcome this shortcoming by using canonical transformation in the context of Hamiltonian formulation. As a deterministic model that describes evolution of sea states without spectral width limitation, Zakharov equation is usually used to reveal some water wave properties such as modulation instability \citep{crawford1981stability} and kinetic equation for spectrum of random wave field \citep{gramstad2013phase}. However, in order to recover the random wave field we should focus on the canonical transformation between the lowest order action variable $b$ and total action variable $A$, i.e., $A(b,b^*)$. In other words, this canonical transformation corresponding bound wave components is helpful to obtain the steady-state solution of random wave field when ignoring dynamic components. \citet{janssen2009some} provided explicit expressions for second-order spectrum, skewness and kurtosis by using canonical transformation in the Hamiltonian theory of water wave. Up to second order in wave steepness, A thorough comparison between the perturbation theory and Hamiltonian theory was carried out by \citet{elfouhaily2000truncated}. They concluded that the coupling coefficients of the surface elevation are identical, which will be confirmed in our paper.

Moreover, up to third order in wave steepness, the wave amplitude dependence in the dispersion relation should be considered to avoid the emergence of secular terms, resulting in a Stokes frequency correction. The third-order dispersion relation of two interacting wave trains was first studied explicitly by \citet{longuet1962phase}. Then, \citet{huang1976dispersion} and \citet{masuda1979dispersion} extended Longuet-Higgins \& Phillips’ analysis to a random wave field in deep water. \citet{taklo2015measurement} and \citet{taklo2017dispersion} confirmed that the dispersion in numerical and experimental random wave field deviates from theoretical linear dispersion relation. Recently, based on the Zakharov equation, \citet{stuhlmeier2019nonlinear} derived the nonlinear dispersion relation for the finite water depth, which is expressed in terms of the energy spectrum.

The object of this paper is to derive a steady-state third-order solution for surface gravity waves in finite water depth, which overcomes the singularity in transfer function and has practicality value. The solution includes explicit expressions for velocity potential, surface elevation and velocity potential at free surface, as well as the nonlinear dispersion relation. We use two approaches: One is the perturbation expansion technique taking the original water wave equation as a starting point and the other is the canonical transformation that corresponds to bound harmonics in the framework of Hamiltonian theory. A comprehensive comparison between these two solutions is carried out. 
 Our solution is an extension of third-order Stokes wave from a single wave to a random wave field, and it is an extension of the work by \citet{sharma1981second} from second-order interactions to third-order interactions. Finally, based on the third-order solution, some consequences are given.

The paper is organized as follows. \S ~\ref{sec:TheorDescri} presents the theoretical framework of surface gravity waves including classic description and Hamiltonian description. The third-order solution for multidirectional irregular waves in finite water depth by using the perturbation expansion technique is derived in \S~\ref{sec:Perturbation}. The third-order solution using the canonical transformation in the context of Hamiltonian theory of water wave is presented in \S~\ref{sec:Canonical transformation}. We comprehensively compare and discuss these two types of approximate analytical solutions in \S~\ref{sec:comparison}. Based on the third-order solution some consequences are given in \S~\ref{sec:consequences}. Moreover, transfer functions in the perturbation solution and coefficients of canonical transformation are provided in Appendix \ref{appA} and Appendix \ref{appB}, respectively.

\section{Statement of problem}\label{sec:TheorDescri}
\subsection{Governing equations}
We adopt a Cartesian coordinate system with the horizontal coordinates ${\bf x} =(x,y) $ on the mean water level (MWL) and the vertical coordinate $z$ pointing upwards, and consider a fluid domain $\mathcal{D}^\zeta_{-h}$ which is vertically bounded by the free surface $\zeta({\bf x},t)$ and the seabed $z=-h$:
\begin{equation*}
\mathcal{D}^\zeta_{-h}=\left\lbrace \left( {\bf x},z\right):{\bf x}\in\mathbb{R}^d, -h\leqslant z\leqslant\zeta({\bf x},z) \right\rbrace,  
\end{equation*}
where $d$ is the horizontal spatial dimension. Under the assumption of irrotational flow in a homogeneous incompressible and inviscid fluid, the motion of the flow field can be described by a velocity potential $\phi({\bf x},z,t)$, which obeys the Laplace equation
\begin{equation}\label{eq:Laplace}
\nabla^2\phi=0,
\end{equation}
subject to the combination of kinematic and dynamic boundary conditions
\begin{equation}\label{eq:SFC}
\phi_{tt}+g\phi_{z}+2\nabla\phi\cdot\nabla\phi_{t}+\frac{1}{2}\nabla\phi\cdot\nabla(\nabla\phi\cdot\nabla\phi)=0 \quad \mbox{on} \quad z=\zeta({\bf x},t),
\end{equation}
where $\nabla=(\partial/\partial x, \partial/\partial y, \partial/\partial z)$ denotes the gradient operator and $g$ is the gravitational acceleration. Vanishing of the normal velocity on the seabed leads to
\begin{equation}\label{eq:bottom}
\phi_{z}=0 \quad \mbox{on}\quad z=-h.
\end{equation}
The surface elevation can be obtained by the dynamic boundary condition
\begin{equation}\label{eq:DFC}
\phi_{t}+\frac{1}{2}\nabla\phi\cdot\nabla\phi+g\zeta=C(t)  \quad \mbox{on}\quad z=\zeta({\bf x},t),
\end{equation}
and the kinematic boundary condition is given by
\begin{equation}\label{eq:KFC}
\zeta_{t}+\nabla_{\bf x}\phi\cdot\nabla_{\bf x}\zeta-\phi_{z}=0 \quad \mbox{on}\quad  z=\zeta({\bf x},t),
\end{equation}
where $\nabla_{\bf x}=(\partial/\partial x, \partial/\partial y)$ denotes the horizontal gradient operator. In addition, the free surface potential $\psi({\bf x},t)$ can be obtained from the following form
\begin{equation}\label{eq:Fpotential}
\psi-\phi=0\quad \mbox{on}\quad z=\zeta({\bf x},t).
\end{equation}

\subsection{Hamiltonian description}
Modern water wave theories start from the pioneering work of \citet{zakharov1968stability}, who discovered the Hamiltonian structure of the potential flow theory with respect to canonically conjugate variables $\zeta({\bf x},t)$ and $\psi({\bf x},t)=\phi\left({\bf x},z=\zeta,t\right)$:
\begin{equation}\label{eq:Hamilton}
\frac{\partial\zeta}{\partial t}=\frac{\delta H}{\delta\psi},\ \ \ \frac{\partial\psi}{\partial t}=-\frac{\delta H}{\delta\zeta},
\end{equation}
where $\delta$ stands for the variational derivative and $H$ is Hamiltonian 
\begin{equation}\label{eq:Hamiltonian}
H={\rm K.E.}+{\rm P.E.}=\frac{1}{2}\int_{\mathbb{R}^{d}}d{\bf x}\int_{-h}^{\zeta}\left(\left|\nabla_{\bf x}\phi\right|^{2}+\phi_{z}^{2}\right)dz+\frac{1}{2}g\int_{\mathbb{R}^{d}}\zeta^{2}d{\bf x},
\end{equation}
which represents the mechanical energy of water waves, consisting of the kinetic K.E. and potential P.E. energies. Obviously, Hamilton equations \eqref{eq:Hamilton} are equivalent to the kinematic (\ref{eq:KFC}) and dynamic conditions (\ref{eq:DFC}). In this paper, we consider a two-dimensional horizontal plane ($d=2$) for multidirectional irregular surface gravity waves.
 
Let us define the two-dimensional Fourier transform conventions:
\begin{equation*}
\hat{f}({\bf k})=\frac{1}{(2\pi)^2}\int{f}({\bf x})e^{-i{\bf k}\cdot{\bf x}}d{\bf x}, \ \ f({\bf x})=\int\hat{f}({\bf k})e^{i{\bf k}\cdot{\bf x}}d{\bf k}.
\end{equation*}

Taking the Fourier transform with respect to the horizontal coordinates, the solution to the Laplace equation (\ref{eq:Laplace}) satisfying the bottom boundary condition (\ref{eq:bottom}) reads
\begin{equation}\label{eq:phi}
\phi({\bf x},z)=\int\hat{\phi}({\bf k})\frac{\cosh{k(z+h)}}{\cosh{kh}}e^{i{\bf k}\cdot{\bf x}}d{\bf k}, \ \ \ \hat{\phi}({\bf k})=\hat{\phi}^{*}(-{\bf k}),
\end{equation}
where ${\bf{k}}$ is the wavevector with the modulus value $k=|{\bf k}|$ and the time-dependence $t$ has been suppressed.

Assuming $\epsilon=|{\bf{k}}|\zeta\ll 1$ and using (\ref{eq:phi}), the Fourier transform of $\psi(\bf x)$ can be expanded in Taylor series up to $\mathcal{O}(\epsilon^3)$
\begin{equation}\label{eq:Fourierpsi}
\hat{\psi}({\bf{k}})=\hat{\phi}({\bf{k}})+\int q_1\hat{\phi}_1\hat{\zeta}_2\delta_{0-1-2}d{\bf{k}}_{12}+\frac{1}{2}\int k_1^2\hat{\phi}_1\hat{\zeta}_2\hat{\zeta}_3\delta_{0-1-2-3}d{\bf{k}}_{123},
\end{equation}
where
\begin{equation}
q({\bf{k}})=|{\bf{k}}|\tanh(|{\bf{k}}|h).
\end{equation} 
By inverting the above relation (\ref{eq:Fourierpsi}) iteratively  relative to $\hat{\phi}$ and proper symmetrization, we get
\begin{equation}\label{eq:Fourierphi}
\hat{\phi}({\bf{k}})=\hat{\psi}({\bf{k}})-\int q_1\hat{\psi}_1\hat{\zeta}_2\delta_{0-1-2}d{\bf{k}}_{1,2}-\int D_{0,1,2,3}^{(3)}\hat{\psi}_1\hat{\zeta}_2\hat{\zeta}_3\delta_{0-1-2-3}d{\bf{k}}_{123}
\end{equation}
where 
\begin{equation}\label{eq:D3}
D_{0,1,2,3}^{(3)}=\frac{1}{2}\left( k_1^2-q_1q_{0-2}-q_1q_{0-3}\right).
\end{equation}
Similarly, the truncated kinetic energy ${\rm K.E.}$ can be acquired by Taylor series expansion:
\begin{align}
{\rm K.E.}=&\frac{1}{2}\int q_0\hat{\phi}_{0}\hat{\phi}_{0}^{*}d{\bf k}+\frac{1}{2}\int (q_0q_1-{\bf{k}}\cdot{\bf{k}}_1)\hat{\phi}_{0}\hat{\phi}_{1}\hat{\zeta}_2 \delta_{0+1+2}d{\bf{k}}_{012}\nonumber\\
+&\frac{1}{2}\int (q_0q_1-{\bf{k}}\cdot{\bf{k}}_1)\frac{(q_0+q_1)}{2}\hat{\phi}_{0}\hat{\phi}_{1}\hat{\zeta}_2\hat{\zeta}_3\delta_{0+1+2+3} d{\bf{k}}_{0123}
\end{align}
Substituting (\ref{eq:Fourierphi}) into the above equation and collecting the same order, correct to $\mathcal{O}(\epsilon^4)$, yield
\begin{align}\label{eq:kenergy}
{\rm K.E.}=&\frac{1}{2}\int q_0\hat{\psi}_{0}\hat{\psi}_{0}^{*}d{\bf k}+\frac{1}{2}\int E_{0,1,2}^{(3)}\hat{\psi}_{0}\hat{\psi}_{1}\hat{\zeta}_2 \delta_{0+1+2}d{\bf{k}}_{012}\nonumber\\
+&\frac{1}{2}\int E_{0,1,2,3}^{(4)}\hat{\psi}_{0}\hat{\psi}_{1}\hat{\zeta}_2\hat{\zeta}_3\delta_{0+1+2+3} d{\bf{k}}_{0123},
\end{align}
where
\begin{subequations}\label{eq:Etranscoe}
	\begin{align}
	E_{0,1,2}^{(3)}= & -\frac{1}{2}\left(q_{0}q_{1}+{\bf k}\cdot{\bf k}_{1}\right),\\
	E_{0,1,2,3}^{(4)}= & -\frac{1}{8}\left(2(k^{2}q_{1}+k_1^{2}q_{0})-q_{0}q_{1}\left(q_{0+2}+q_{0+3}+q_{1+2}+q_{1+3}\right)\right).
	\end{align}
\end{subequations}
The kinetic energy ${\rm K.E.}$ is expressed in terms of $\hat{\zeta}$ and $\hat{\psi}$ while the potential energy ${\rm P.E.}$ is only related to $\hat{\zeta}$. These cumbrous calculations was carried out by \cite{krasitskii1994reduced} up to the fifth-order terms inclusive. 

One of the important advantages of Hamiltonian formalism is that it allows us to use a wide class of canonical transformations for different purposes and maintain the Hamiltonian structure. For the above water wave problem, the canonical transformations are used to completely remove the bound wave component. However, the present study focuses on the bound parts. We will give the  steady-state third-order solutions in \S ~\ref{sec:Canonical transformation} by means of canonical transformation. 
\section{Perturbation expansion}\label{sec:Perturbation}
A classic perturbation series approach, also known as the Stokes expansion, is usually used to solve the boundary-value problem of water wave. The approximate analytical solutions can be expressed by a sum of the first few terms:
\begin{subequations}\label{eq:perturbation}
	\begin{align}
	\zeta&=\epsilon\zeta_1+\epsilon^2\zeta_2+\epsilon^3\zeta_3+\cdots \nonumber\\
	\phi&=\epsilon\phi_1+\epsilon^2\phi_2+\epsilon^3\phi_3+\cdots \nonumber\\
	\psi&=\epsilon\psi_1+\epsilon^2\psi_2+\epsilon^3\psi_3+\cdots \nonumber
	\end{align}	
\end{subequations}
where $\epsilon\ll 1$ is a small parameter that measures the wave steepness.

To eliminate the secular terms at third order, we introduce the new time-scale $\tau$ by the transformation, correct to $\mathcal{O}(\epsilon^2)$
\begin{equation}\label{eq:tau}
\tau=t\left[ 1+\epsilon^2\mu+\mathcal{O}(\epsilon^4)\right], 
\end{equation}
where $\mu$ is a nonlinear coefficient to be determined.
Notice that the straining of time coordinate is equivalent to the procedure in which one has $\Omega\;t=\omega\;\tau$ and a frequency change is introduced by
\begin{equation}\label{eq:nonlinear0}
\Omega=\omega\left[ 1+\epsilon^2\mu+\mathcal{O}(\epsilon^4)\right].
\end{equation}
Here the relation between the nonlinear frequency $\Omega$ and the linear frequency $\omega$ constitutes the third-order dispersion relation which will be determined in (\ref{eq:nonlinearDis}). Taking (\ref{eq:tau}), the time derivative is replaced by
\begin{equation}\label{eq:Stime}
\partial_t = \left(  1+\epsilon^2\mu \right)  \partial_{\tau}.
\end{equation}
Then, we expand the free-surface condition (\ref{eq:SFC}), (\ref{eq:DFC}) and (\ref{eq:Fpotential}) in Taylor series with respect to the MWL and substitute (\ref{eq:perturbation}) and (\ref{eq:Stime}) into the Taylor series expansion. Finally, we collect the same order terms to obtain
\begin{subequations}\label{eq:BVP}
	\begin{alignat}{2}
	\nabla^2\phi_i &= 0, \qquad \qquad &-h<z<0,\\
	\phi_{i,z} &=0,    &z=-h,\\
	\phi_{i,\tau\tau}+g\phi_{i,z}&= f_i^{(1)}, &z=0,\\
	\phi_{i,\tau}+g\zeta_i&= f_i^{(2)},  &z=0,\\
	\psi_i-\phi_i &= f_i^{(3)},  &z=0,
	\end{alignat}	
\end{subequations}
where $f_i^{(m)}, m=1,2,3$ are forcing functions in terms of lower-order quantities and $\epsilon$ has been drawn into physical variables. Starting from the first equations at $i=1$, which are linear homogeneous and solved easily, the following equations can be solved successively by the lower-order solution. 

\subsection{The first-order solution}\label{subsec:1Psolution}
At the first order, $i=1$, (\ref{eq:BVP}) is homogeneous and the first-order solution including \textit{N} wave components can be written as
\begin{subequations}\label{eq:1Psol}
	\begin{alignat}{1}
	\zeta_1&=\sum_{n}a_n\cos{\theta_n},\label{subeq:1Psur}\\
	\phi_1&=\sum_{n}\frac{g}{\omega_{n}}\frac{\cosh{{k}_n(z+h)}}{\cosh k_nh}a_n\sin{\theta_n},\\
	\psi_1&=\sum_{n}\frac{g}{\omega_{n}} a_n\sin{\theta_n},
	\end{alignat}	
\end{subequations}
where the phase function is given by
\begin{equation} \label{eq:totalphase}
\theta_n={\bf{k}}_n\cdot{\bf{x}}-\omega_{n} \tau+\varepsilon_n\quad\mbox{or}\quad \theta_n={\bf{k}}_n\cdot{\bf{x}}-\Omega_{n} t+\varepsilon_n,
\end{equation}
with wavenumber $k_n=|{\bf k}_n|$, $\varepsilon_n$ is uniformly distributed in the interval $[0,2\pi)$ and $a_n$ is the wave amplitude of individual wave component with the linear angular frequency $\omega_{n}$ and the wavevector ${\bf{k}}_n=(k_{nx},k_{ny})$. They satisfy the linear dispersion relation
\begin{equation}\label{eq:LDR}
\omega^2=gk\tanh kh.
\end{equation} 
The first-order solution is also known as the linear random wave model representing a sum of Airy waves with different frequencies and directions of propagation.

\subsection{The second-order solution}\label{subsec:2Psolution}
At the second-order, $i=2$, taking gravity as the only restoring force in surface gravity waves leads to no resonance in the second-order solution, and the inhomogeneous terms $f_2^{(m)}, m=1,2,3$ read 
\begin{subequations}\label{eq:secondforce}
	\begin{alignat}{1}
	f_2^{(1)}&=-\zeta_1 (\phi_{1,\tau\tau z}+g\phi_{1,zz})-2\nabla\phi_1\cdot\nabla\phi_{1,\tau},\\
	f_2^{(2)}&=-\zeta_1\phi_{1,\tau z}-\frac{1}{2}\nabla\phi_1\cdot\nabla\phi_1,\\
	f_2^{(3)}&=\;\;\;\zeta_1\phi_{1,z}.
	\end{alignat}
\end{subequations}
Substituting $\phi_1, \zeta_{1}$ from \eqref{eq:1Psol} into \eqref{eq:secondforce}, we obtain the second-order solution
\begin{subequations}\label{eq:2Psol}
	\begin{alignat}{1}
	\zeta_2&=\sum_{n,m}{E}_{n\pm m}a_n a_m\cos{\theta_{n\pm m}}+C_1,\label{eq:secondsolusur}\\ 
	\phi_2&=\sum_{n,m}{F}_{n\pm m}\frac{\cosh{k_{n\pm m}(z+h)}}{\cosh k_{n\pm m}h}a_n a_m\sin{\theta_{n\pm m}}+{\bf U}\cdot{\bf{x}}+C_2 \; t,\\
	\psi_2&=\sum_{n,m}{G}_{n\pm m} a_n a_m \sin{\theta_{n\pm m}}+{\bf U}\cdot{\bf{x}}+C_2 \; t,
	\end{alignat}
\end{subequations}
where 
\begin{gather*}
\theta_{n\pm m}=\theta_n \pm \theta_m,\\
k_{n\pm m}=|{\bf{k}}_n\pm {\bf{k}}_m|=\sqrt{(k_{nx}\pm k_{mx})^2+(k_{ny}\pm k_{my})^2}.
\end{gather*}
Since the coordinate system has been set at the MWL, i.e., $\bar{\zeta}=0$,  the constant $C_1$ can be determined
\begin{equation}
	C_1=-\sum_{n}E_{n-n}a_n^2.
\end{equation}
  The vector ${\bf U}$ corresponds to induced mean flow,  and the constant $C_2$ gives a contribution to the mean pressure which is therefore related to the mean surface. 
  
  There are two wave celerity definitions proposed by \cite{Stokes47} to determine the uniform flow ${\bf U}$. According to the Stokes' first wave celerity definition, namely, the mean value of the horizontal Eulerian flow velocity equal to zero ($\overline{\nabla_{\bf x}\phi}=0$). Thus guaranteeing the periodicity of velocity potential, we obtain
\begin{equation}
-\sum_{n}F_{n-n}\theta_{n-n}a_n^2={\bf U}\cdot{\bf{x}}+C_2 \; t
\end{equation}
In the above two equations, $E_{n-n}$ and $F_{n-n}\theta_{n- n}$ are two non-unique limits. The discussion about them will be presented in \S~\ref{subsec:non-uniqueness}. Alternatively, the vector ${\bf U}$ can be determined by the Stokes' second definition of wave celerity, some details will be given in \S~\ref{subsec:volume flux}.

In \eqref{eq:2Psol}, ${E}_{n\pm m}, {F}_{n\pm m}, {G}_{n\pm m}$ are second-order transfer functions for the surface elevation, velocity potential and free-surface potential respectively, with the following forms
\begin{subequations}
	\begin{alignat}{1}
	{E}_{n\pm m}&=\frac{\omega_{n\pm m}}{g}F_{n\pm m}+\frac{\mathcal{B}_{n\pm m}}{g}\\
	{F}_{n\pm m}&=-\frac{\mathcal{A}_{n\pm m}}{\beta_{n\pm m}}\\
	{G}_{n\pm m}&={F}_{n\pm m}+\mathcal{C}_{n\pm m}
	\end{alignat}
\end{subequations}
where 
\begin{gather}
\omega_{n\pm m}=\omega_{n}\pm\omega_{m},\notag\\
\beta_{n\pm m}=\omega_{n\pm m}^2-gk_{n\pm m}\tanh{k_{n\pm m}h},\label{eq:Resonance2}
\end{gather}
and $\mathcal{A}_{n\pm m}, \mathcal{B}_{n\pm m}$ and $\mathcal{C}_{n\pm m}$ are second-order transfer coefficients given in Appendix \ref{appA}.

The second-order solutions are consistent with the original
derivation by \citet{sharma1981second}. Moreover, it is worth noting that the second-order transfer functions have the following relationship:

\begin{equation}\label{eq:symmetry2}
{E}_{n\pm m}= {E}_{m\pm n};\quad {F}_{n\pm m}= \pm {F}_{m\pm n}; \quad {G}_{n\pm m}= \pm {G}_{m\pm n}
\end{equation}

\subsection{The third-order solution}
At the third-order, $i=3$, the amplitude dispersion appears and the corresponding inhomogeneous terms $f_3^{(m)}, m=1,2,3$ read
\begin{subequations}\label{eq:thirdforce}
	\begin{alignat}{1}
	f_3^{(1)}=&-2\mu\phi_{1,\tau\tau}-\frac{1}{2}\zeta_1^2(\phi_{1,\tau\tau zz}+g\phi_{1,zzz})-\frac{1}{2}\nabla\phi_1\cdot\nabla(\nabla\phi_1\cdot\nabla\phi_1)\notag\\
	&-2\zeta_1\left[ \nabla\phi_1\cdot\nabla\phi_{1,\tau}\right]_z-2\nabla\phi_1\cdot\nabla\phi_{2,\tau}-2\nabla\phi_2\cdot\nabla\phi_{1,\tau} \notag \\[3pt]
	&-\zeta_1(\phi_{2,\tau\tau z}+g\phi_{2,zz})-\zeta_2(\phi_{1,\tau\tau z}+g\phi_{1,zz})\\ \label{subeq:thirdforcePhi}
	f_3^{(2)}=&-\mu\phi_{1,\tau}-\frac{1}{2}\zeta_1^2\phi_{1,\tau zz}-\zeta_1\phi_{2,\tau z}-\zeta_2\phi_{1,\tau z}-\nabla\phi_1\cdot\nabla\phi_2-\zeta_1\left[ \nabla\phi_1\cdot\nabla\phi_1\right]_z \\ \label{subeq:thirdforcesur}
	f_3^{(2)}=&\quad\;\zeta_1\phi_{2,z}+\zeta_2\phi_{1,z}+\frac{1}{2}\zeta_1^2\phi_{1,zz}
	\end{alignat}
\end{subequations}

\subsubsection{Nonlinear dispersion relation}
Upon substituting the first-order solutions \eqref{eq:1Psol} and second-order solutions \eqref{eq:2Psol} into \eqref{subeq:thirdforcePhi}, we obtain
\begin{align}\label{eq:thirdPhi}
\frac{\partial^2\phi_3}{\partial\tau^2}+g\frac{\partial\phi_3}{\partial z}&=2g\sum_{n}\mu_n\omega_{n}a_n\sin{\theta_n} +\sum_{n,m,p}\mathcal{A}_{n\pm m\pm p}a_n a_m a_p\sin{\theta_{n\pm m\pm p}}\notag\\
+&\sum_{n,m}\left\lbrace 2gF_{m-m}{\bf k}_n\cdot ({\bf k}_m-{\bf k}_m)-E_{m-m}\left( {\omega_n^3}-\frac{g^2k_n^2}{\omega_n}\right)\right\rbrace a_na_m^2\sin\theta_n
\end{align}
where $\mathcal{A}_{n\pm m\pm p}$ is the third-order transfer coefficient given in Appendix \ref{appA}.

The unknown coefficient $\mu_n$ is now chosen in such a way that no $\sin\theta_{n}$ terms remain and by (\ref{eq:nonlinear0}) we obtain the nonlinear angular frequency $\Omega$ with the following form

\begin{equation}\label{eq:nonlinearD}
\varOmega_n=\omega_n(1+\mu_n)=\omega_n\left( 1+\sum_{m}{e}_{nm}{W}_{nm}a_m^2\right) 
\end{equation}
where $e_{nm}=1/2$ for $n=m$ and $e_{nm}=1$ for $n\neq m$, and
\begin{align}\label{eq:FrequencyT}
{W}_{nm}({\bf{k}}_n,{\bf{k}}_m)=&\frac{1}{4}\left( \frac{2\omega_{m}^2+\omega_{n}^2}{\omega_{n}\omega_{m}}{\bf{k}}_n\cdot{\bf{k}}_m+ k_m^2\right) \notag\\
&-\frac{1}{2}(E_{n+m}+E_{n-m})\left( \frac{\omega_{m}^2}{g}-\frac{g}{\omega_{n}\omega_{m}}{\bf{k}}_n\cdot{\bf{k}}_m\right)  \notag\\
&-\frac{1}{2}\frac{\omega_{n}}{g}\left[  F_{n+m}k_{n+m}\tanh(k_{n+m}h)+F_{n-m}k_{n-m}\tanh(k_{n-m}h)\right] \notag\\
&+\frac{1}{2}\frac{F_{n+m}}{\omega_{n}\omega_{m}}\left[ (\omega_{n}-\omega_{m})({\bf{k}}_n\cdot{\bf{k}}_m+k_m^2)+\omega_{m}k_{n+m}^2\right]  \notag\\
&+\frac{1}{2}\frac{F_{n-m}}{\omega_{n}\omega_{m}}\left[ (\omega_{n}+\omega_{m})({\bf{k}}_n\cdot{\bf{k}}_m-k_m^2)+\omega_{m}k_{n-m}^2\right] 
\end{align}
Considering the wavenumber energy spectrum $S({\bf{k}})$ contained in the interval from ${\bf{k}}$ to ${\bf{k}}+d{\bf{k}}$, we obtain
\begin{equation}\label{eq:atoS}
\sum_{{\bf{k}}}^{{\bf{k}}+d{\bf{k}}}\frac{1}{2}a_n^2=S({\bf{k}})d{\bf{k}}.
\end{equation}
Using (\ref{eq:nonlinearD}) and (\ref{eq:atoS}) and taking the limit $d{\bf{k}}\rightarrow 0$, (\ref{eq:nonlinearD}) is rewritten as
\begin{equation}\label{eq:nonlinearDis}
\varOmega({\bf{k}})=\omega({\bf{k}})\left( 1+\int e_{{\bf{k}},{\bf{k}}_1}{W}({\bf{k}},{\bf{k}}_1)S({\bf{k}}_1)d{\bf{k}}_1\right), 
\end{equation}
where
\begin{equation*}  
e_{{\bf{k}},{\bf{k}}_1}=\left\{  
\begin{array}{lr}  
1, &   {\bf{k}}={\bf{k}}_1 \\  
2,   &   \mbox{ otherwise} 
\end{array}  
\right.  
\end{equation*}

\subsubsection{Third-order transfer functions}
After determining the coefficient $\mu_n$, the third-order solutions become
\begin{subequations}\label{eq:thirdsolu}
	\begin{alignat}{1}
	\zeta_3&=\sum_{n,m,p}E_{n\pm m\pm p}a_n a_m a_p\cos{\theta_{n\pm m\pm p}},\\ \label{eq:thirdsolusur}
	\phi_3&=\sum_{n,m,p}F_{n\pm m\pm p}\frac{\cosh{k_{n\pm m\pm p}(z+h)}}{\cosh {k_{n\pm m\pm p}h}}a_n a_m a_p\sin{\theta_{n\pm m\pm p}},\\	
	\psi_3&=\sum_{n,m,p}G_{n\pm m\pm p} a_n a_m a_p \sin{\theta_{n\pm m\pm p}}, 
	\end{alignat}
\end{subequations}
where 
\begin{gather*}
\theta_{n\pm m\pm p}=\theta_n \pm \theta_m\pm \theta_p, \\
k_{n\pm m\pm p}=|{\bf{k}}_n\pm {\bf{k}}_m\pm {\bf{k}}_p|=\sqrt{(k_{nx}\pm k_{mx}\pm k_{px})^2+(k_{ny}\pm k_{my}\pm k_{py})^2}.
\end{gather*}
In (\ref{eq:thirdsolu}), $E_{n\pm m\pm p}, F_{n\pm m\pm p}, G_{n\pm m\pm p}$ are third-order transfer functions for the surface elevation, velocity potential and free-surface potential respectively. Their expressions are as follows:
\begin{subequations}\label{eq:thirdtran}
	\begin{alignat}{1}
	E_{n\pm m\pm p}&=\frac{\omega_{n\pm m\pm p}}{g}F_{n\pm m\pm p}+\frac{\mathcal{B}_{n\pm m\pm p}}{g}\\
	F_{n\pm m\pm p}&=-\frac{\mathcal{A}_{n\pm m\pm p}}{\beta_{n\pm m\pm p}}\\
	G_{n\pm m\pm p}&=F_{n\pm m\pm p}+\mathcal{C}_{n\pm m\pm p}
	\end{alignat}
\end{subequations}
where
\begin{gather}
  \omega_{n\pm m\pm p}=\omega_{n}\pm\omega_{m}\pm\omega_{p},\notag\\
  \beta_{n\pm m\pm p}=\omega_{n\pm m\pm p}^2-gk_{n\pm m\pm p}\tanh(k_{n\pm m\pm p}h),\label{subeq:Resonance}
\end{gather}
and $\mathcal{B}_{n\pm m\pm p}$ and $\mathcal{C}_{n\pm m\pm p}$ are third-order transfer coefficients given in Appendix \ref{appA}. Similar to (\ref{eq:symmetry2}), the coefficients in third-order solutions should be symmetrized. For example, $E_{n+ m+ p}$ is invariant for interchanging the indices $(n, m, p)$ and $E_{n+ m- p}$ is invariant for interchanging the indices $(n, m)$.

\subsubsection{Third-order correction to the linear terms}\label{subsubsec:3to1term}

Similar to the Stokes-type correction to linear frequency $\omega$, there are third-order corrections corresponding to other linear terms, termed as quasi-linear terms. Their transfer functions can be obtained by dropping terms involving $E_{n-n}$ and $F_{n-n}$ in the third-order transfer function , but the transfer function of quasi-linear velocity potential which is forced to be zero to remove secular terms, namely
\begin{equation}
F_{nm}=\sum_{m}e_{nm}{Q}_{nm}\equiv 0.
\end{equation}
The transfer functions of quasi-linear surface elevation and free-surface velocity potential are given by
\begin{equation}
E_{nm}=\sum_{m}e_{nm}{P}_{nm},\quad G_{nm}=\sum_{m}e_{nm}{R}_{nm},
\end{equation}
respectively, where
\begin{align}
P_{nm}({\bf{k}}_n,{\bf{k}}_m)&=\frac{1}{4}\left( k_n^2-k_m^2-\frac{\omega_{n}}{\omega_{m}}{\bf{k}}_n\cdot{\bf{k}}_m\right) \notag\\
&\quad+\frac{1}{2}(E_{n+m}+E_{n-m})\left( \frac{\omega_{m}^2}{g}+\frac{g}{\omega_{n}\omega_{m}}{\bf{k}}_n\cdot{\bf{k}}_m\right) \notag\\
&\quad+\frac{1}{2}\frac{\omega_{n}}{g}\left[ F_{n+m}k_{n+m}\tanh{k_{n+m}h}+F_{n-m}k_{n-m}\tanh{k_{n-m}h}\right] \notag\\
&\quad+\frac{1}{2}\frac{F_{n+m}}{\omega_{n}\omega_{m}}\left[ \omega_{m}k_{n+m}^2-(\omega_{n}+\omega_{m})({\bf{k}}_n\cdot{\bf{k}}_m+k_m^2)\right] \notag\\
&\quad+\frac{1}{2}\frac{F_{n-m}}{\omega_{n}\omega_{m}}\left[
\omega_{m}k_{n-m}^2-(\omega_{n}-\omega_{m})({\bf{k}}_n\cdot{\bf{k}}_m-k_m^2)\right],
\end{align}
and
\begin{align}
  R_{nm}({\bf{k}}_n,{\bf{k}}_m)&=\frac{1}{4}\frac{gk_n^2}{\omega_n}+\omega_m\left(E_{n-m}- E_{n+m}\right)\notag\\
  &\quad+F_{n+m}k_{n+m}\tanh{k_{n+m}h}+F_{n-m}k_{n-m}\tanh{k_{n-m}h}.
\end{align}

We have derived asymptotic solutions for surface gravity waves up to third order. In comparison with the third-order solutions derived by \cite{madsen2012third}, there are some differences in the third-order part (\ref{eq:thirdsolu}). This is because different approaches are used to remove the secular terms. We give a correction to the first-order surface elevation by specifying the coefficient of $\sin{\theta_{n}}$ terms from velocity potential to be zero, while  Madsen $\&$ Fuhrman chose to correct the first-order velocity potential. From the third-order transfer functions such as $F_{n+m-p}$, it is clear that the perturbation solution will breakdown when $\beta_{n\pm m\pm p}$ \eqref{subeq:Resonance} approaches zero. Thus, the perturbation solution is of little help for practical applications. 
 
\section{Canonical transformation}\label{sec:Canonical transformation}
 Drawing on \citet{zakharov1968stability}, we introduce the action variable $A({\bf{k}})$ and its complex conjugate by the first canonical transformation:
\begin{equation}\label{eq:lineartransform}
\hat{\zeta}({\bf{k}})={N}({\bf{k}})(A({\bf{k}})+A^*(-{\bf{k}})),\ \  \hat{\psi}({\bf{k}})=-i{M}({\bf{k}})(A({\bf{k}})-A^*(-{\bf{k}}))
\end{equation}
where $i$ denotes the imaginary unit and 
\begin{equation}\label{eq:NM}
{N}({\bf{k}})=\sqrt{\frac{\omega({\bf{k}})}{2g}},\quad {M}({\bf{k}})=\sqrt{\frac{g}{2\omega({\bf{k}})}}.
\end{equation}
Then, the Hamiltonian of water wave system can be written as a series expansion in integer powers of $A$ and $A^*$, and the Hamilton equations (\ref{eq:Hamilton}) reduce to a single equation
\begin{equation}\label{eq:HamiltonA}
	i\frac{\partial A({\bf k})}{\partial t}=\frac{\delta H}{\delta A^*({\bf k})}
\end{equation}
Substituting the canonical transformation (\ref{eq:lineartransform}) into \eqref{eq:HamiltonA} gives
\begin{align}\label{eq:evolutionA}
\frac{\partial A}{\partial t}+i\omega A =&-i\int\left\lbrace U_{0,1,2}^{(1)}A_1 A_2 \delta_{0-1-2}+2U_{2,1,0}^{(1)}A_1^* A_2 \delta_{0+1-2}+U_{0,1,2}^{(3)}A_1^* A_2^* \delta_{0+1+2}\right\rbrace d{\bf{k}}_{12}\nonumber\\
&-i\int\left\{ V_{0,1,2,3}^{(1)}A_1 A_2 A_3 \delta_{0-1-2-3}+V_{0,1,2,3}^{(2)}A_1^* A_2 A_3 \delta_{0+1-2-3}\right.\nonumber\\
&\quad\ \ \left.+3V_{3,2,1,0}^{(1)}A_1^* A_2^* A_3 \delta_{0+1+2-3}+V_{0,1,2,3}^{(4)}A_1^* A_2^* A_3^*\delta_{0+1+2+3}\right\} d{\bf{k}}_{123},
\end{align}
which is the canonical formulation of the truncated irrotational Euler equation up to third-order terms.

However, the evolution equation (\ref{eq:evolutionA}) is not optimal and contains the dynamic and bound wave components. To remove the bound harmonics, the second canonical transformation $A(b,b^*)$ is introduced to obtain a new evolution equation for $b$, namely the so-called Zakharov equation: 

\begin{equation}\label{eq:Zakharov}
\frac{\partial b}{\partial t}+i\omega b=-i\int T_{0,1,2,3} b_1^* b_2 b_3 \delta_{0+1-2-3}d{\bf{k}}_{123},
\end{equation}
Following \cite{krasitskii1994reduced}, we postulate the canonical transformation $A(b,b^*)$ in the form of integer powers series
\begin{align}\label{eq:Canonicaltransformation}
  A=b&+\int \left\{ A_{0,1,2}^{(1)}b_1 b_2\delta_{0-1-2}+ A_{0,1,2}^{(2)}b_1^{*} b_2\delta_{0+1-2}+A_{0,1,2}^{(3)}b_1^* b_2^*\delta_{0+1+2} \right\} d{\bf{k}}_{12} \nonumber\\
  &+\int \left\{  B_{0,1,2,3}^{(1)}b_1 b_2 b_3\delta_{0-1-2-3}+B_{0,1,2,3}^{(2)}b_1^* b_2 b_3\delta_{0+1-2-3}\right.\nonumber\\
  & \qquad\left. +B_{0,1,2,3}^{(3)}b_1^* b_2^* b_3\delta_{0+1+2-3}+B_{0,1,2,3}^{(4)}b_1^* b_2^* b_3^* \delta_{0+1+2+3}\right\}d{\bf{k}}_{123}
\end{align}
where the non-resonant coefficients $(A^{(1)},A^{(2)},A^{(3)},B^{(1)},B^{(3)},B^{(4)})$ are determined immediately by removing the non-resonant third- and fourth-order contributions in the Hamiltonian $H$, while the resonant coefficient $B^{(2)}$ and Zakharov kernel $T_{0,1,2,3}$ require special treatments due to the presence of quartet resonance. Insisting that $H$ remains real and conservative, all the canonical transformation coefficients and Zakharov kernel should satisfy the natural symmetry. For example, Zakharov kernel $T_{0,1,2,3}$ satisfies $T_{0,1,2,3}=T_{1,0,2,3}=T_{0,1,3,2}=T_{3,2,1,0}$. The derivation of these coefficients are given in Appendix \ref{appB}.

\subsection{Nonlinear dispersion relation}
Zakharov equation clearly shows that the random wave field is a dynamic evolution system subject to quartet resonance. It means that the steady-state solution can be obtained by dropping dynamic components. The initial state is set to obey the linear wave theory, namely, using $b_n=|b_n|\exp(i \arg{b_n})$ the initial Gaussian sea surface (lowest order) is given by
\begin{equation}\label{eq:discreteSUR1}
	\zeta_1({\bf{x}},t)=\sum_{n}{N}({\bf{k}}_n)(|b_n| { e}^{i{\bf{k}}_n\cdot{\bf{x}}+\arg b_n}+c.c) 
\end{equation}
where $c.c$ is complex conjugation. Compared with the first-order surface elevation in the perturbation solution \eqref{subeq:1Psur}, one can easily obtain the following relationship 
\begin{subequations}\label{eq:cVSaOmegaFull}
	\begin{equation}\label{subeq:aVSb}
		a_n = 2{N}_n|b_n|,
	\end{equation}
	\begin{equation}\label{subeq:OmegaFull}
		\Omega_n = -\frac{d}{dt}(\arg b_n),
	\end{equation}
\end{subequations}
where $N_n$ has been defined in \eqref{eq:NM}.

We rewrite the Zakharov equation \eqref{eq:Zakharov} in discrete form
\begin{equation}
	\frac{\partial b_n}{\partial t}+i\omega_n b_n=-i\sum_{m,p,q}T_{nmpq}b_m^* b_p b_q \delta_{n+m-p-q}.
\end{equation}
Separating the real and imaginary part of the above equation yield
\begin{subequations} \label{eq:disZakharov-K} 
	\begin{equation}
		\frac{d|b_n|}{dt}=\sum_{m,p,q}T_{nmpq}\delta_{nmpq}|b_m||b_p||b_q|\sin{\theta_{nmpq}}
	\end{equation}
	\begin{equation}
		\frac{d\arg b_n}{dt}=-\omega_n-|b_n|^{-1}\sum_{m,p,q}T_{nmpq}\delta_{nmpq}|b_m||b_p||b_q|\cos{\theta_{nmpq}}
	\end{equation}
\end{subequations}
where
\begin{equation*}
	\theta_{nmpq}=-\arg b_n-\arg b_m+\arg b_p+\arg b_q
\end{equation*} 

According to \citet{phillips1960dynamics}, the exact quartet resonant criterion satisfies
\begin{equation}\label{eq:ResonanceH}
	{\bf{k}}_n+{\bf{k}}_m={\bf{k}}_p+{\bf{k}}_q,\quad \omega_n+\omega_m=\omega_p+\omega_q,
\end{equation}
Usually, solutions to (\ref{eq:ResonanceH}) can be divided into two categories: trivial and non-trivial. Trivial solutions is easy to find, i.e., (${\bf{k}}_n,{\bf{k}}_m,{\bf{k}}_n,{\bf{k}}_m$). According to the two types of solutions, we rewrite (\ref{eq:disZakharov-K}) as
\begin{subequations}\label{eq:disZakharov-K-1}
	\begin{equation}\label{subeq:disZakharov-K-a}
		\frac{d|b_n|}{dt}=\sum_{m,p\neq n,q\neq n}T_{nmpq}\delta_{nmpq}|b_m||b_p||b_q|\sin{\theta_{nmpq}},
	\end{equation}
	\begin{equation}\label{subeq:disZakharov-K-b}
		\frac{d\arg b_n}{dt}=-\omega_n-\sum_{m}e_{nm}T_{nmnm}|b_m|^2-\sum_{m,p\neq n,q\neq n}T_{nmpq}\delta_{nmpq}|b_m||b_p||b_q|\cos{\theta_{nmpq}},
	\end{equation}
\end{subequations}
where $e_{nm}=1$ for $n=m$ and $e_{nm}=2$ for $n\neq m$. The last terms on the right-hand side of (\ref{eq:disZakharov-K-1}) correspond to non-trivial solution representing resonant quartets. From the above equations, it is shown that trivial components lead to third-order frequency correction while non-trivial components result in the time-dependent wave amplitude. This separation operation allows us to easily remove dynamic components. Removing these quartet resonant terms and integrating over time give
\begin{subequations}
	\begin{equation}\label{subeq:binvariant}
		|b_n(t)|=|b_n(0)|,
	\end{equation}
	\begin{equation}\label{subeq:phaseb}
		\arg b_n=-(\omega_n+\sum_{m}e_{nm}T_{nmnm}|b_m|^2)\;t+\arg(b_n(0)).
	\end{equation}
\end{subequations}
Without the resonant term, it means that wave amplitudes is time-independent and there is no energy exchange between the wave harmonics. $\arg(b_n(0))$ is the initial random phase and equal to $\varepsilon_n$ in \eqref{eq:totalphase}. So \eqref{eq:discreteSUR1} is equivalent to the first-order surface elevation \eqref{subeq:1Psur}.
Substituting (\ref{subeq:phaseb}) into (\ref{subeq:OmegaFull}) gives the discrete third-order dispersion relation
\begin{equation}\label{eq:disNDR-Zakharov}
	\Omega_n=\omega_n+\sum_{m}e_{nm}T_{nmnm}|b_m|^2.
\end{equation}
Using (\ref{eq:atoS}) and (\ref{subeq:aVSb}), \eqref{eq:disNDR-Zakharov} becomes
\begin{equation}\label{eq:cNDR-Zakharov}
	\Omega({\bf{k}})=\omega({\bf{k}})+g\int e_{{\bf{k}},{\bf{k}}_1}T({\bf{k}},{\bf{k}}_1,{\bf{k}},{\bf{k}}_1)\frac{S({\bf{k}}_1)}{\omega({\bf{k}}_1)}d{\bf{k}}_1
\end{equation}   
Recently, \cite{stuhlmeier2019nonlinear} provided the same result in (2.14b) of their work with only a difference of the numerical factor $4\pi^2$. This difference is due to the definition of the Fourier transform. To facilitate comparison with the nonlinear dispersion relation in the perturbation solution, \eqref{eq:cNDR-Zakharov} is rewritten as
\begin{equation}
	\varOmega({\bf{k}})=\omega({\bf{k}})\left( 1+\int e_{{\bf{k}},{\bf{k}}_1}\mathcal{W}({\bf{k}},{\bf{k}}_1)S({\bf{k}}_1)d{\bf{k}}_1\right), 
\end{equation}
where
\begin{equation*}
	\mathcal{W}_{nm}({\bf k}_n,{\bf k}_m)=\frac{g}{\omega_n\omega_m}T({\bf k}_n,{\bf k}_m,{\bf k}_n,{\bf k}_m).
\end{equation*}

Moreover, we offer an alternative to derive the nonlinear dispersion relation. Substituting the canonical transformation (\ref{eq:Canonicaltransformation}) into (\ref{eq:evolutionA}), using $b({\bf{k}},t)=B({\bf{k}},t)\exp{\left( -i\omega({\bf{k}})t\right) }$ and collecting the result up to the third-order terms, one can obtain the following equation
\begin{alignat}{1}\label{eq:midevolutionA}
	\frac{\partial b}{\partial t}+i\omega b =&-i\int \left\{ \left( \varDelta_{0-1-2}A_{0,1,2}^{(1)}+U_{0,1,2}^{(1)}\right) b_1 b_2\delta_{0-1-2}\right.\notag\\
	&\qquad+\left( \varDelta_{0+1-2}A_{0,1,2}^{(2)}+2U_{2,1,0}^{(1)}\right) b_1^* b_2\delta_{0+1-2}\notag\\
	&\;\qquad\left.+\left( \varDelta_{0+1-2}A_{0,1,2}^{(2)}+U_{0,1,2}^{(3)}\right) b_1^* b_2^*\delta_{0+1-2}\right\} d{\bf{k}}_{12}\notag\\
	&-i\int\left\{ \left( \varDelta_{0-1-2-3}B_{0,1,2,3}^{(1)}+Z^{(1)}_{0,1,2,3}+V^{(1)}_{0,1,2,3}\right) b_1b_2b_3 \delta_{0-1-2-3}\right.\notag\\
	&\qquad+\left( \varDelta_{0+1-2-3}B_{0,1,2,3}^{(2)}+Z^{(2)}_{0,1,2,3}+V^{(2)}_{0,1,2,3}\right) b_1^*b_2b_3 \delta_{0+1-2-3}\notag\\
	&\qquad+\left( \varDelta_{0+1+2-3}B_{0,1,2,3}^{(3)}+Z^{(3)}_{0,1,2,3}+3V^{(1)}_{3,2,1,0}\right) b_1^*b_2^*b_3 \delta_{0+1+2-3}\notag\\
	&\qquad\left.+\left( \varDelta_{0+1+2+3}B_{0,1,2,3}^{(4)}+Z^{(4)}_{0,1,2,3}+V^{(4)}_{0,1,2,3}\right) b_1^*b_2^*b_3^* \delta_{0+1+2+3}\right\} d{\bf{k}}_{123}
\end{alignat}
where the coefficients $Z^{()}$ are given in Appendix \ref{appB}. Then, the nonlinear frequency $\Omega_n$ can be obtained by
\begin{equation}
	\Omega=-\operatorname {Im}\overline{\left[ \frac{\dot{b}(t)}{b(t)}\right] }
\end{equation}
where overline denotes time-averaging. Only the lowest order terms are preserved to get 
\begin{equation}
	\Omega_{n}=\omega_{n}+\sum_{m} e_{nm} \widetilde{T}_{n,m,n,m}|b_m|^2,
\end{equation}
where the subscript $(0,1)$ is replaced by $(n,m)$ and
\[
\widetilde{T}_{n,m,n,m}=\frac{1}{2}\left( Z^{(2)}_{n,m,n,m}+Z^{(2)}_{n,m,m,n}\right) +V^{(2)}_{n,m,n,m}.
\] 
Obviously, it is straightforward to see that $\widetilde{T}_{n,m,n,m}={T}_{n,m,n,m}$. Therefore, we can get the equivalent results for the nonlinear dispersion relation whether starting from the evolution equation (\ref{eq:evolutionA}) or the Zakharov equation (\ref{eq:Zakharov}).

\subsection{The Hamiltonian solution}
\subsubsection{Surface elevation and free-surface velocity potential}
In order to present the solution in Fourier series, the second canonical transformation \eqref{eq:Canonicaltransformation} is rewritten in a discrete form
\begin{align}\label{eq:disCanonicaltransformation}
	A=&\sum_n b_n\delta_{0-n}+\sum_{n,m} \left\{ A_{0,n,m}^{(1)}b_n b_m\delta_{0-n-m}+A_{0,n,m}^{(3)}b_n^* b_m^*\delta_{0+n+m}\right. \notag\\
	&+\sum_{n,m,p} \left\{  B_{0,n,m,p}^{(1)}b_n b_m b_p\delta_{0-n-m-p}+B_{0,n,m,p}^{(2)}b_n^* b_m b_p\delta_{0+n-m-p}\right.\notag\\
	& \quad\qquad\left. +B_{0,n,m,p}^{(3)}b_n^* b_m^* b_p\delta_{0+n+m-p}+B_{0,n,m,p}^{(4)}b_n^* b_m^* b_p^* \delta_{0+n+m+p}\right\}.
\end{align}
Here, we have replaced subscripts $(1,2,3)$ by $(n,m,p)$ to make a more intuitive comparison with the perturbation solution.
Note that there is no singularity in the transfer coefficient $B_{0,1,2,3}^{(2)}$ given in \eqref{eq:B2}. This ensures that the Hamiltonian solution does not collapse in the vicinity of the resonance point.

Since the coordinates are fixed on the MWL, the ensemble average of surface elevation should be zero. Thus, we shall subtract the average value generated by the quadratic part in the canonical transformation \eqref{eq:disCanonicaltransformation}. This average is responsible for the induced mean current and mean surface. After eliminating the average, we rewrite the canonical transformation in a simple form
\begin{equation}\label{eq:simpleCanonicaltrans}
A=\epsilon b+\epsilon^2 \tilde{p}(b,b^*)+\epsilon^3 q(b,b^*),
\end{equation}
where $\tilde{p}=p-\bar{p}\delta_{1-2}$ is the quadratic part with zero mean, $p$ corresponds to the quadratic part and $q$ is the cubic part.

By substituting the above equation into the first canonical transformation \eqref{eq:lineartransform}, using the properties of delta function:
\begin{equation*}
	\delta({\bf k})=\frac{1}{\left(2\pi\right)^2}\int e^{-i{\bf k}\cdot{\bf x}}d{\bf x}, \ \ \int\delta({\bf k})e^{i{\bf k}\cdot{\bf x}}d{\bf k}=1,
\end{equation*} 
and then applying the inverse Fourier transform, we recover the surface elevation 
\begin{align}\label{eq:surface}
\zeta=&\sum_n a_n\cos\theta_n +\sum_{n,m} \mathcal{E}_{n\pm m}a_na_m \cos\theta_{n\pm m}+C_1\notag\\
&+\sum_{n,m,p}\left\lbrace \mathcal{E}_{n+ m+p}a_na_ma_p \cos\theta_{n+m+p}+\mathcal{E}_{n\pm m-p}a_na_ma_p \cos\theta_{n\pm m-p} \right\rbrace 
\end{align}
and the free-surface velocity potential
\begin{align}\label{eq:surphi}
\psi=&\sum_n \frac{g}{\omega_n} a_n \sin\theta_n+\sum_{n,m} \mathcal{G}_{n\pm m}a_na_m \sin\theta_{n+m}+{\bf U}\cdot{\bf{x}}+C_2\;t\notag\\
&+\sum_{n,m,p} \left\lbrace \mathcal{G}_{n+m+p}a_na_ma_p \sin\theta_{n+m+p}+\mathcal{G}_{n\pm m-p}a_na_ma_p  \sin\theta_{n+m-p}\right\rbrace,
\end{align}
where the vector ${\bf U}$ and the coefficient $C_1, C_2$ have been determined in the second-order perturbation solution \eqref{eq:2Psol}. The nonlinear transfer functions for $\zeta$ and $\psi$ have the following explicit expressions: \\
Second-order
\begin{subequations}
	\begin{equation}
	\mathcal{E}_{n+m}=\frac{1}{2}{N}_{n+m}{N}^{-1}_{n}{N}^{-1}_{m}\left( A_{n+m,n,m}^{(1)}+A_{-n-m,n,m}^{(3)}\right),
	\end{equation}
	\begin{equation}
	\mathcal{E}_{n-m}=-\frac{1}{2}{N}_{n-m}{N}^{-1}_{n}{N}^{-1}_{m} \left( A_{m,n,-n+m}^{(1)}+A_{n,m,-m+n}^{(1)}\right),
	\end{equation}
	\begin{equation}
	\mathcal{G}_{n+m}=\frac{1}{2}{M}_{n+m}{N}^{-1}_{n}{N}^{-1}_{m}\left( A_{n+m,n,m}^{(1)}-A_{-n-m,n,m}^{(3)}\right),
	\end{equation}	
	\begin{equation}
	\mathcal{G}_{n-m}=\frac{1}{2}{M}_{n-m}{N}^{-1}_{n}{N}^{-1}_{m} \left( A_{m,n,-n+m}^{(1)}-A_{n,m,-m+n}^{(1)}\right);
	\end{equation}
\end{subequations}
Third-order
\begin{subequations}
	\begin{equation}
	\mathcal{E}_{n+m+p}=\frac{1}{4}{N}_{n+m+p}{N}^{-1}_{n}{N}^{-1}_{m}{N}^{-1}_{p}\left( B_{n+m+p,n,m,p}^{(1)}+B_{-n-m-p,n,m,p}^{(4)}\right),
	\end{equation}
	\begin{equation}
	\mathcal{G}_{n+m+p}=\frac{1}{4}{M}_{n+m+p}{N}^{-1}_{n}{N}^{-1}_{m}{N}^{-1}_{p}\left( B_{n+m+p,n,m,p}^{(1)}-B_{-n-m-p,n,m,p}^{(4)}\right),
	\end{equation}
	\begin{equation}
	\mathcal{E}_{n+m-p}=\frac{1}{4}{N}_{n+m-p}{N}^{-1}_{n}{N}^{-1}_{m}{N}^{-1}_{p} B_{-n-m+p,n,m,p}^{(3)},
	\end{equation}
	\begin{equation}
	\mathcal{G}_{n+m-p}=-\frac{1}{4}{M}_{n+m-p}{N}^{-1}_{n}{N}^{-1}_{m}{N}^{-1}_{p} B_{-n-m+p,n,m,p}^{(3)},
	\end{equation}
	\begin{equation}
	\mathcal{E}_{n-m-p}=\frac{1}{4}{N}_{n-m-p}{N}^{-1}_{n}{N}^{-1}_{m}{N}^{-1}_{p} B_{-n+m+p,n,m,p}^{(2)},
	\end{equation}		
	\begin{equation}
	\mathcal{G}_{n-m-p}=-\frac{1}{4}{M}_{n-m-p}{N}^{-1}_{n}{N}^{-1}_{m}{N}^{-1}_{p} B_{-n+m+p,n,m,p}^{(2)}.
	\end{equation}
\end{subequations}

\subsubsection{Velocity potential}
After determining the surface elevation \eqref{eq:surface} and the free-surface velocity potential \eqref{eq:surphi}, the velocity potential can be established by the relationship \eqref{eq:Fourierphi} between the potential at the free surface and at the MWL. Using the first canonical transformation \eqref{eq:lineartransform} one find
\begin{equation}\label{eq:psixi}
\hat{\psi}_1\hat{\zeta}_2=-iM_1N_2\left\lbrace A_1A_2+A_{1}A_{-2}^*-c.c(1\leftrightarrow-1,2\leftrightarrow-2)\right\rbrace
\end{equation}
and
\begin{align}\label{eq:psixixi}
\hat{\psi}_1\hat{\zeta}_2\hat{\zeta}_3=&-iM_1N_2N_3\left\{ A_1A_2A_3+A_1A_2A^{*}_{-3}+A_1A^*_{-2}A_3+A_1A_{-2}^*A_{-3}^* \right.\notag\\
&\left.\qquad\qquad-c.c(1\leftrightarrow-1,2\leftrightarrow-2,3\leftrightarrow-3)\right\}
\end{align}
Substituting the second canonical transformation \eqref{eq:simpleCanonicaltrans} into \eqref{eq:psixi} and \eqref{eq:psixixi}, then collecting the result up to third order in $\epsilon$, give 
\begin{align}\label{eq:psixi0}
\hat{\psi}_1\hat{\zeta}_2=&-iM_1N_2\left\{ \epsilon^2(b_1b_2+b_{1}b_{-2}^*)+\epsilon^3 (b_1\tilde{p}_2+\tilde{p}_1b_2+b_{1}\tilde{p}_{-2}^*+\tilde{p}_{1}b_{-2}^*)\right.\notag\\
&\qquad\qquad\left.-c.c(1\leftrightarrow-1,2\leftrightarrow-2) \right\}
\end{align}
and 
\begin{align}\label{eq:psixixi0}
\hat{\psi}_1\hat{\zeta}_2\hat{\zeta}_3=&-iM_1N_2N_3\left\{ \epsilon^3(b_1b_2b_3+b_1b_2b^*_{-3}+b_1b^*_{-2}b_3+b_1b^*_{-2}b^*_{-3})\right.\notag\\
&\qquad\qquad\qquad\left.-c.c(1\leftrightarrow-1,2\leftrightarrow-2,3\leftrightarrow-3)\right\}.
\end{align}
Substituting \eqref{eq:psixi0} and \eqref{eq:psixixi0} into the relation \eqref{eq:Fourierphi}, we obtain
\begin{align}\label{eq:phivspsi}
	\hat{\phi}({\bf k})-\hat{\psi}({\bf k})=&i\int \left\lbrace \frac{1}{2}\left( q_1M_1N_2+q_2M_2N_1\right)b_1b_2\delta_{0-1-2}+q_2M_2N_1 b_1^*b_2\delta_{0+1-2}\right\rbrace d{\bf k}_{12} \notag\\
  	&+i\int\left\lbrace  C_{0,1,2,3}^{(1)}b_1b_2b_3\delta_{0-1-2-3}+C_{0,1,2,3}^{(2)}b_1^*b_2b_3\delta_{0+1-2-3}\right.  \notag\\
	       &\left. \;\;\;\;\;\;+ C_{0,1,2,3}^{(3)}b_1^*b_2^*b_3\delta_{0+1+2-3}+C_{0,1,2,3}^{(4)}b_1^*b_2^*b_3^*\delta_{0+1+2+3}\right\rbrace d{\bf k}_{123} \notag\\
	 &-c.c(1\rightarrow-1,2\rightarrow-2,3\rightarrow-3),
\end{align}
where the coefficients $C^{()}$ are given in the Appendix \ref{appB} with the appropriate symmetrization. Then, substituting the above equation into (\ref{eq:phi}) and using the inverse Fourier transform, we recover the velocity potential
\begin{align}\label{eq:phiH}
\phi=&\sum_n \frac{g}{\omega_n} \frac{\cosh{k_n(z+h)}}{\cosh{k_nh}}a_n \sin\theta_n+\sum_{n,m}\mathcal{F}_{n\pm m}\frac{\cosh{k_{n\pm m}(z+h)}}{\cosh{k_{n\pm m}h}}a_na_m \sin\theta_{n\pm m}\notag\\
&+\sum_{n,m,p} \mathcal{F}_{n+m+p}\frac{\cosh{k_{n+m+p}(z+h)}}{\cosh{k_{n+m+p}h}}a_na_ma_p \sin\theta_{n+m+p}\notag\\
&+\sum_{n,m,p} \mathcal{F}_{n\pm m-p}\frac{\cosh{k_{n\pm m-p}(z+h)}}{\cosh{k_{n\pm m-p}h}}a_na_ma_p \sin\theta_{n\pm m-p}+{\bf U}\cdot{\bf{x}}+C_2\;t,
\end{align}
where
\begin{subequations}
	\begin{align}
	\mathcal{F}_{n\pm m}&=\mathcal{G}_{n\pm m}-{X}_{n\pm m},\\
	\mathcal{F}_{n+m+p}&=\mathcal{G}_{n+m+p}-{X}_{n+m+p},\\
	\mathcal{F}_{n\pm m-p}&=\mathcal{G}_{n\pm m-p}-{X}_{n\pm m-p}.
	\end{align}
\end{subequations}
Here, the expression of transfer coefficients ${C}$ are given by 
\begin{subequations}
	\begin{align}
	{X}_{n\pm m}&=\frac{1}{4}\left( \omega_{n}\pm\omega_m\right)\\
	{X}_{n+m+p}&=\frac{1}{4}{N}^{-1}_{n}{N}^{-1}_{m}{N}^{-1}_{p}\left( C_{n+m+p,n,m,p}^{(1)}-C_{-n-m-p,n,m}^{(4)}\right),\\
	{X}_{n+m-p}&=-\frac{1}{4}{N}^{-1}_{n}{N}^{-1}_{m}{N}^{-1}_{p}C_{-n-m+p,n,m,p}^{(3)}\\
	{X}_{n-m-p}&=-\frac{1}{4}{N}^{-1}_{n}{N}^{-1}_{m}{N}^{-1}_{p}C_{-n+m+p,n,m,p}^{(2)}  
	\end{align}
\end{subequations}

\subsection{Transfer functions of quasi-linear terms}
There are also quasi-linear terms which are third order in the Hamiltonian solution. Unlike the perturbation expansion theory in which the quasi-linear term of velocity potential is forced to be zero, all quasi-linear terms in the Hamiltonian solution exist and have different transfer functions. For the quasi-linear terms, the above transfer functions cannot be used directly and some changes take place, due to the subtraction of mean value in the second canonical transformation 

The transfer functions of quasi-linear surface elevation and free-surface velocity potential are given by
\begin{equation}
	\mathcal{E}_{nm}=\sum_{m}e_{nm}\mathcal{P}_{nm},\quad \mathcal{G}_{nm}=\sum_{m}e_{nm}\mathcal{R}_{nm},
\end{equation}
where
\begin{align}
	\mathcal{P}_{nm}=&\frac{1}{2}{N}^{-2}_{m}\bigg[  A_{n,m,-n-m}^{(3)}A_{m,n,-m-n}^{(3)}+A_{m,n,m-n}^{(1)}A_{m,n,m-n}^{(1)}\nonumber\\
	&\quad -A_{n+m,n,m}^{(1)}A_{m+n,m,n}^{(1)}-A_{n,m,n-m}^{(1)}A_{n,m,n-m}^{(1)}\nonumber\\
	&\quad-\omega_{n}^{-1}\big(  U_{-n,m,-n-m}^{(1)}A_{-n-m,n,m}^{(3)}-U_{-n+m,-n,m}^{(1)}A_{m,n,m-n}^{(1)}  \nonumber\\
	&\quad+U_{m,-n,m+n}^{(1)}A_{n+m,n,m}^{(1)}- U_{-n,m,n-m}^{(3)}A_{n,m,n-m}^{(1)}+\frac{3}{2}V_{m,m,n,-n}^{(1)}\big)\bigg],
\end{align}
\begin{align}
	\mathcal{R}_{nm}=&-\frac{1}{2}M_nN_n^{-1}{N}^{-2}_{m}\bigg[  -A_{n,m,-n-m}^{(3)}A_{m,n,-m-n}^{(3)}-A_{m,n,m-n}^{(1)}A_{m,n,m-n}^{(1)}\nonumber\\
	&\quad +A_{n+m,n,m}^{(1)}A_{m+n,m,n}^{(1)}+A_{n,m,n-m}^{(1)}A_{n,m,n-m}^{(1)}\nonumber\\
	&\quad-\omega_{n}^{-1}\big(  U_{-n,m,-n-m}^{(1)}A_{-n-m,n,m}^{(3)}-U_{-n+m,-n,m}^{(1)}A_{m,n,m-n}^{(1)}  \nonumber\\
	&\quad+U_{m,-n,m+n}^{(1)}A_{n+m,n,m}^{(1)}- U_{-n,m,n-m}^{(3)}A_{n,m,n-m}^{(1)}+\frac{3}{2}V_{m,m,n,-n}^{(1)}\big)\bigg].
\end{align}
The transfer function of quasi-linear velocity potential in Hamiltonian solutions is given by
\begin{equation}
	\mathcal{F}_{nm}=\sum_{m}e_{nm}\mathcal{Q}_{nm}
\end{equation}
where
\begin{align}
\mathcal{Q}_{nm}=&\mathcal{R}_{nm}+\omega_m\mathcal{E}_{n+m}-q_{n+m}\mathcal{G}_{n+m}\notag\\
	&+\frac{1}{2}\frac{g}{\omega_m}\left( D^{(3)}_{-n,m,-m,-n}-D^{(3)}_{n,m,n,-m}\right)-\frac{1}{2}\frac{g}{\omega_n}D^{(3)}_{n,n,m,-m},
\end{align}
where $D^{(3)}_{0,1,2,3}$ has given in \eqref{eq:D3}.
Similar to (\ref{eq:nonlinearDis}), the sum of linear and quasi-linear terms in Hamiltonian theory also can be written as
\begin{subequations}\label{eq:quasi-surH}
	\begin{align}
		\zeta_{1}^s&=\zeta_1\left( 1+\int e_{{\bf{k}},{\bf{k}}_1}\mathcal{P}({\bf{k}},{\bf{k}}_1)S({\bf{k}}_1)d{\bf{k}}_1\right), \\
		\psi_{1}^s&=\psi_1\left( 1+\frac{\omega({\bf k})}{g}\int e_{{\bf{k}},{\bf{k}}_1}\mathcal{R}({\bf{k}},{\bf{k}}_1)S({\bf{k}}_1)d{\bf{k}}_1\right), \\
		\phi_{1}^s&=\phi_1\left( 1+\frac{\omega({\bf k})}{g}\int e_{{\bf{k}},{\bf{k}}_1}\mathcal{Q}({\bf{k}},{\bf{k}}_1)S({\bf{k}}_1)d{\bf{k}}_1\right), 
	\end{align}	
\end{subequations}

\section{Comparison and discussion}\label{sec:comparison}
\subsection{A simple example involving trichromatic interactions }
Now we have derived the third-order asymptotic solutions by not only the singular perturbation method from the original water wave equations (hereafter referred to as Perturbation solution), but also by the canonical transformation in the Hamiltonian formalism (hereafter referred to as Hamiltonian solution). In order to compare the two solutions, we consider an elementary trichromatic interaction defined by
\begin{equation*}
{\bf{k}}_n=(1.48667,0.85833),\quad {\bf{k}}_m=(1.74536,-0.63526),\quad {\bf{k}}_p=(2.02010,\varepsilon),
\end{equation*}
where $\varepsilon$ controls the distance from the resonance point (\ref{eq:ResonanceH}). The amplitudes are   
$$a_n=0.025\ {\rm m},\quad a_m=0.025\ {\rm m},\quad a_p=0.050\ {\rm m}$$
with $h=1.0 \ {\rm m}$, $g=9.81\ {\rm m\ s^{-2}}$. The corresponding values of steepness are $\epsilon_n=0.043,\epsilon_m=0.047$ and $\epsilon_p=k_p a_p$ is related to $\varepsilon$. All initial phases are taken to be zero. Based on the above third-order analytical solutions, we will analyze the surface elevation $\zeta$ and the velocity field $\nabla\phi=(u,v,w)$.

In fact, only two primary harmonics are needed to produce a quartet resonance when a certain harmonic is computed twice. Thus, the above trichromatic case provides various combinations that are likely to satisfy the resonance criterion. To measure the distance from the resonance point, we introduce a ratio defined by
\begin{equation*}
\alpha_{n\pm m\pm p}=\frac{\min |\varDelta_{nmpq}/\omega_{i}|}{\max \epsilon_i^2}\qquad i=n,m,p
\end{equation*}
A larger value of $\alpha$ means a larger distance from the resonance point. $\varDelta_{nmpq}=\omega_{n}+\omega_{m}-\omega_{p}-\omega_{q}$ denotes the angular frequency mismatch in which $\omega_{q}$ is a third-order daughter-wave frequency and calculated by the linear dispersion relation with the wavevector ${\bf k}_n+{\bf k}_m-{\bf k}_p$. The value of $\varepsilon$ in wavevector ${\bf k}_p$ allows us to control the distance from the quartet resonance point. We consider two cases with different distances from the resonance point, corresponding to $\varepsilon=0$ and $\varepsilon=0.95$ respectively, to compare the performance of the two kinds of solutions. For case $\varepsilon=0$, the parameters of the above trichromatic interaction are the same as those in \cite{madsen2012third}. For case $\varepsilon=0.95$, the combination $(n,m,p)$ is very close to exact resonance.

\begin{figure}
	\centering
	\subfigure[Case:~$\varepsilon=0$]{
		\label{subfig:surface0}
		\includegraphics[width=0.85\textwidth]{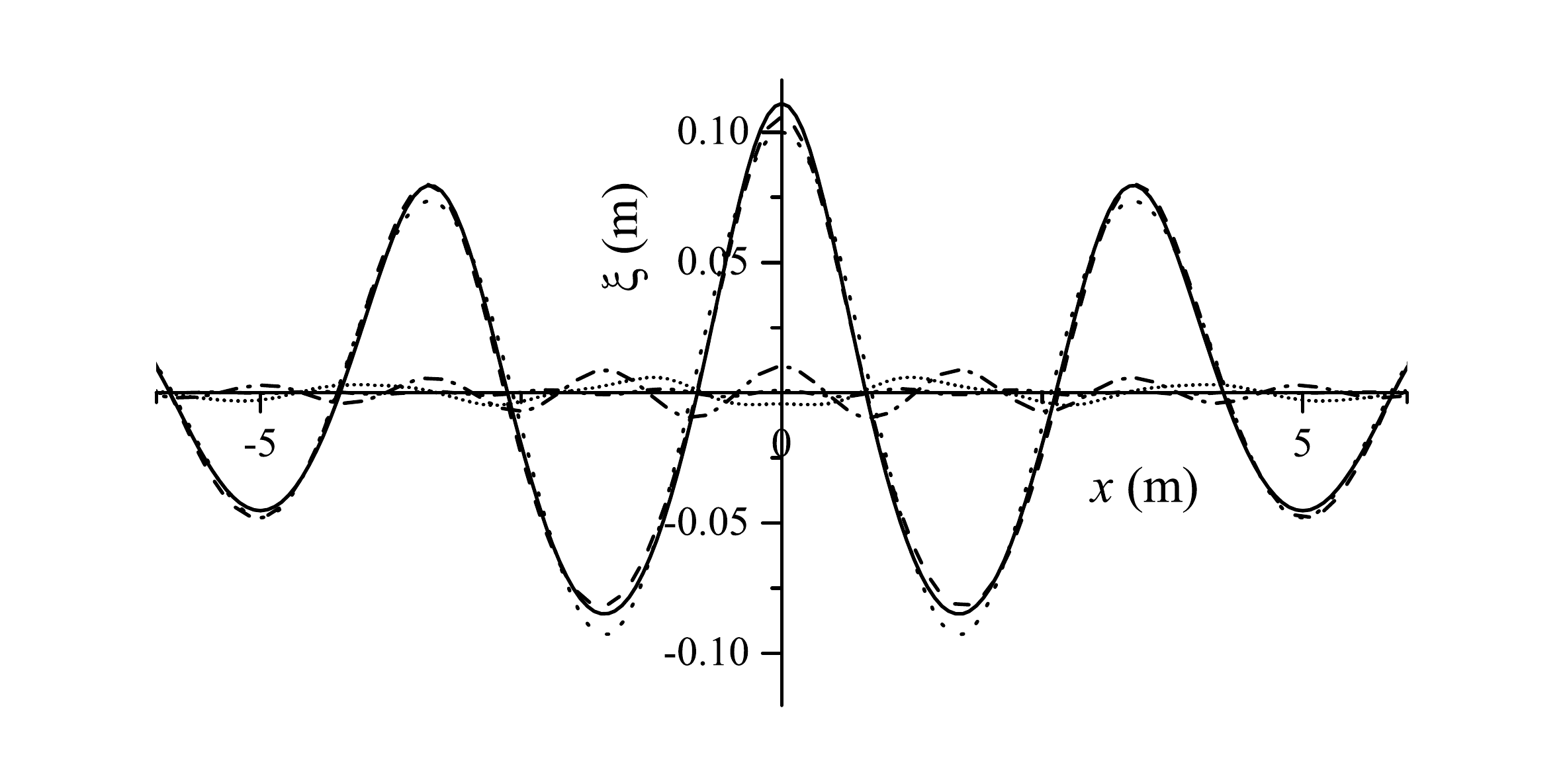} 
    }
	\quad
	\subfigure[Case:~$\varepsilon=0.95$]{
		\label{subfig:surface0.95}
		\includegraphics[width=0.85\textwidth]{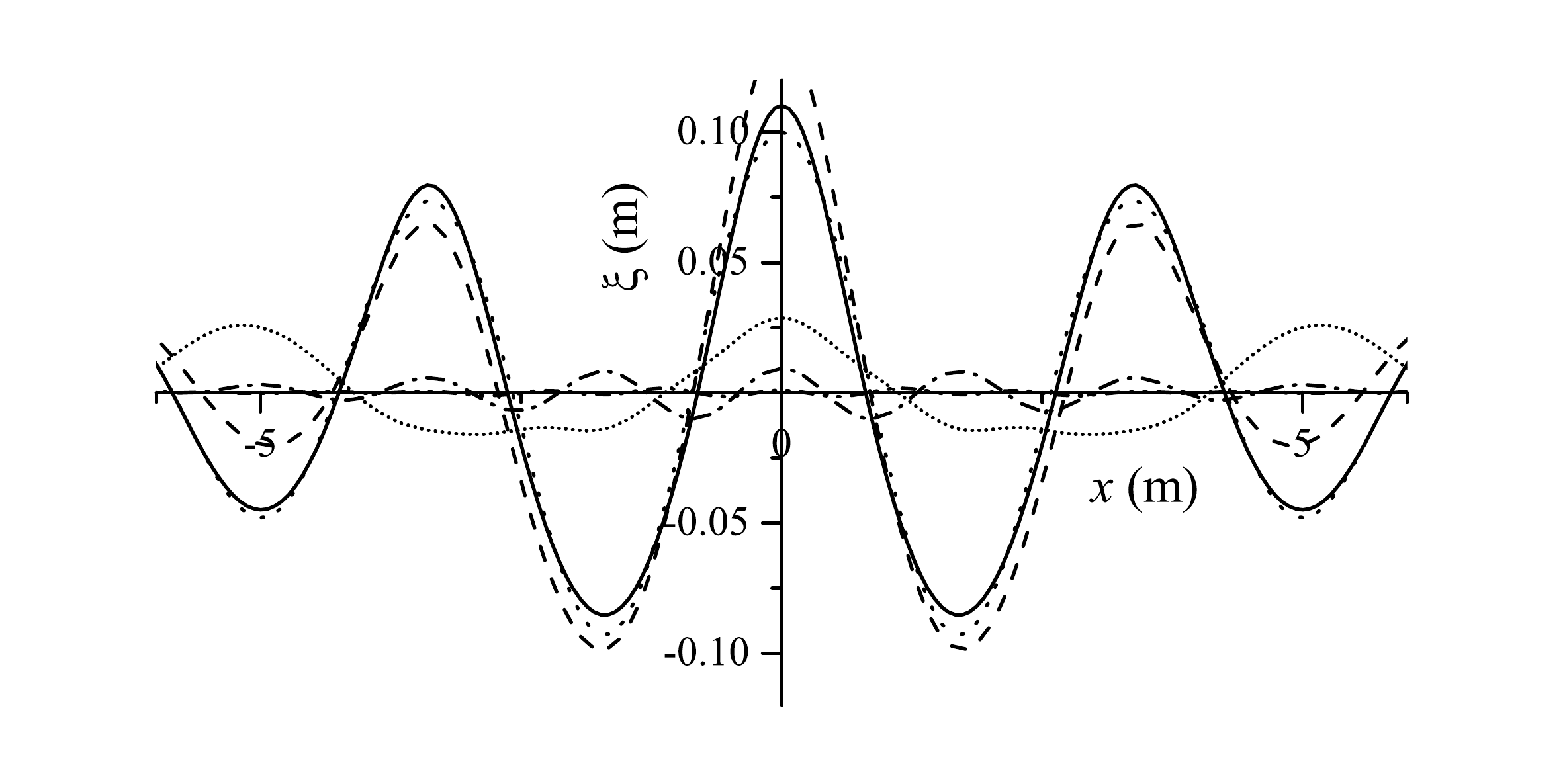}	
    }   
	\caption{ Surface elevation $\zeta=\zeta_1+\zeta_2+\zeta_3$ along the centreline $y=0$ at $t=0$:
		\protect\tikz[baseline]{\protect\draw[line width=0.5mm] (0,.8ex)--++(1,0) ;}~$\zeta^H$,
		\protect\tikz[baseline]{\protect\draw[line width=0.5mm,dashed] (0,.8ex)--++(1,0) ;}~$\zeta^P$,
		\protect\tikz[baseline]{\protect\draw[line width=0.5mm,loosely dotted] (0,.8ex)--++(1,0);}~$\zeta_1$,
		\protect\tikz[baseline]{\protect\draw[line width=0.5mm,dash dot] (0,.8ex)--++(1,0) ;}~$\zeta_2$, 
		\protect\tikz[baseline]{\protect\draw[line width=0.5mm,dash dot dot] (0,.8ex)--++(1,0) ;}~$\zeta_3^H$, 
		\protect\tikz[baseline]{\protect\draw[line width=0.5mm,densely dotted] (0,.8ex)--++(1,0) ;}~$\zeta_3^P$. 
		The superscript $H$ refers to the Hamiltonian solution while $P$ refers to the perturbation solution.}
	\label{fig:surface}
\end{figure}

\begin{figure}
	\centering
	\subfigure[Case:~$\varepsilon=0$]{
			\includegraphics[width=0.85\textwidth]{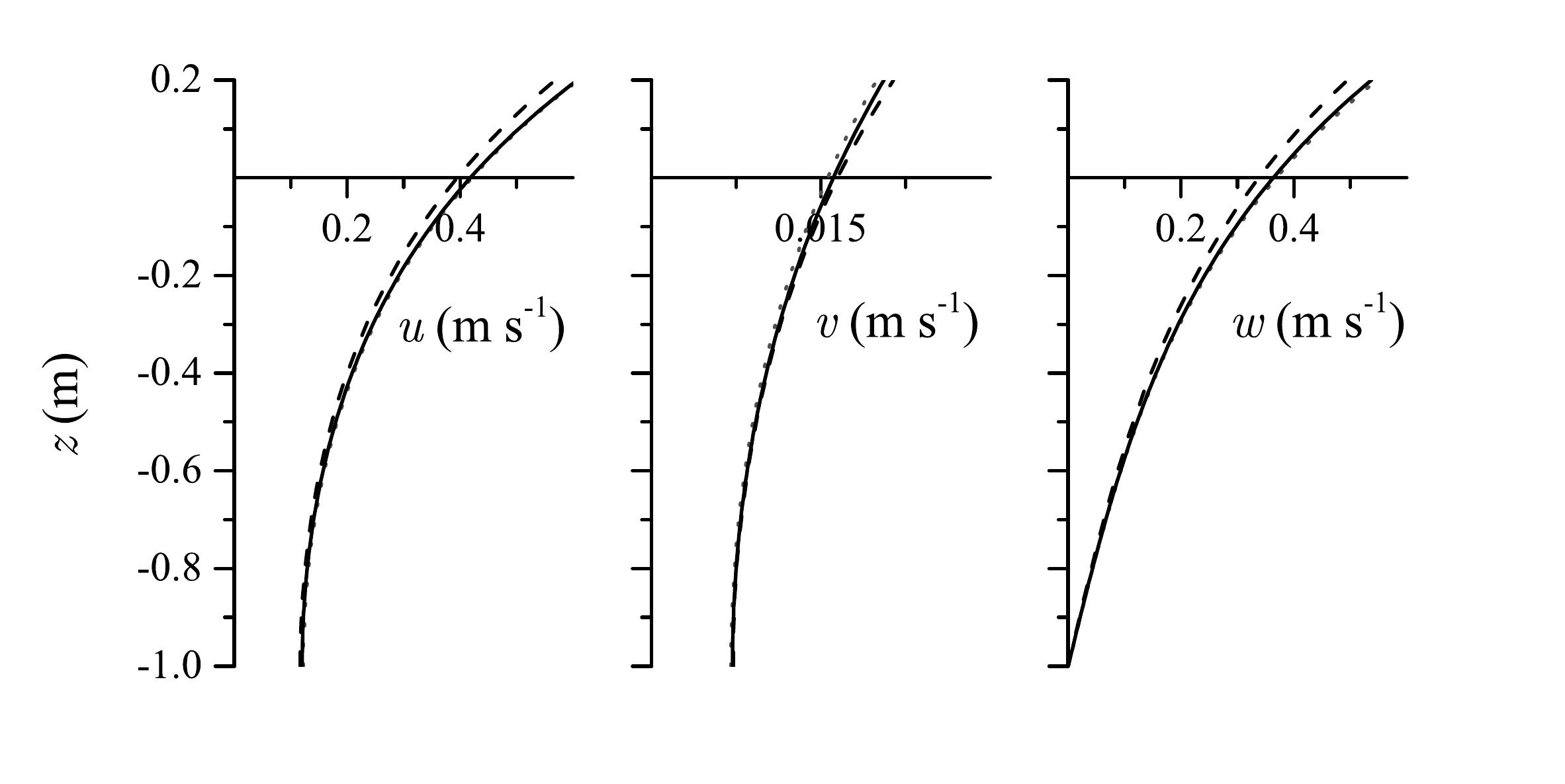} 
	}
\quad
	\subfigure[Case:~$\varepsilon=0.95$]{
			\includegraphics[width=0.85\textwidth]{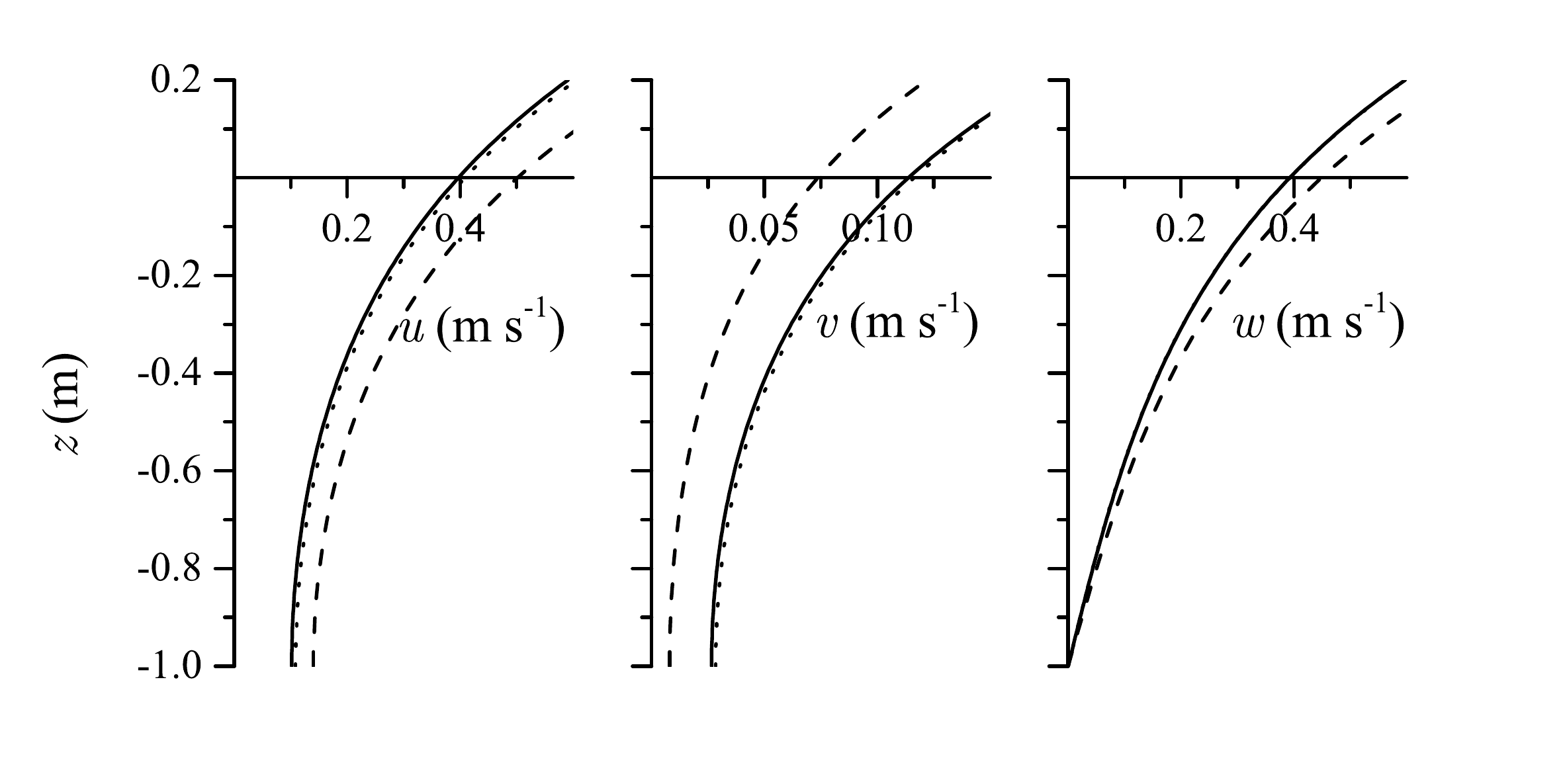}
	}
   \caption{Velocity profile $(u,v,w)$ at the central point $(x, y)=(0,0)$:
   	\protect\tikz[baseline]{\protect\draw[line width=0.5mm] (0,.8ex)--++(1,0) ;}~$(u,v,w)^H$,
   	\protect\tikz[baseline]{\protect\draw[line width=0.5mm,dashed] (0,.8ex)--++(1,0) ;}~$(u,v,w)^P$,
   	\protect\tikz[baseline]{\protect\draw[line width=0.5mm,loosely dotted] (0,.8ex)--++(1,0) ;}~$(u,v,w)_1$. The superscript $H$ refers to the Hamiltonian solution while $P$ refers to the perturbation solution.}
   \label{fig:velocity}
\end{figure}

Figures~\ref{fig:surface} and~\ref{fig:velocity} show the variation of surface elevation and velocity of the trichromatic example, respectively. The upper diagrams of both figures correspond to the case of $\varepsilon=0$, in which the trichromatic example has a finite distance from resonance point. In this case, the smallest value of $\alpha$ is 4.70 corresponding to the combination $(m,p,p)$. The lower diagrams of both figures correspond to the case of $\varepsilon=0.95$ in which some combinations are very close to satisfying the resonance criterion, especially the combination $(n,m,p)$  with $\alpha_{n+ m- p}=0.07$. Overall, according to the $\alpha$ values listed in the last column of Tables~\ref{tab:TF123a} and~\ref{tab:TF123b}, the case $\varepsilon=0.95$ has more combinations close to the resonance point.

As shown in Figure~\ref{fig:surface}, the surface elevation is plotted against the abscissa $x$ along the centreline $(y=0)$. The first to third order components of surface elevation $(\zeta_1,\zeta_2,\zeta_3)$ are also presented in this figure. Both $\zeta_1$ and $\zeta_2$ from the two analytical solutions are coincident; the same conclusion also appears in the velocity field, although the second-order part is not given in Figure~\ref{fig:velocity}. Therefore, the different results given by the two theoretical solutions are only attributed to the difference in third-order part. For the perturbation solutions, in case $\varepsilon=0$ the magnitude of $\zeta_3^P$ is slightly smaller than that of $\xi_2$, because most combinations are far away from the resonance point. However, as the combinations approaching the resonance point increase in case $\varepsilon=0.95$, the magnitude of $\zeta_3^P$ far exceeds that of $\xi_2$. Such results are incorrect and violate the assumption of perturbation expansion method (i.e, $\zeta_1\gg\zeta_2\gg\zeta_3$). Figure~\ref{fig:velocity} shows the velocity profile at the central point $(x,y)=(0,0)$. It is obviously seen that the velocity field in the perturbation solution has a larger deviation from the benchmark (i.e, ${(u,v,w)}_1$) than those in the Hamiltonian solution, and the degree of deviation is even more pronounced in the case $\varepsilon=0.95$. In contrast to the perturbed solution, the Hamiltonian solution seems to be more reasonable. On the basis of our analysis of the simple trichromatic example, it can be concluded that  when describing the steady-state resonant waves the perturbation solutions collapse and give unreliable results, but Hamiltonian solutions overcome this shortcoming due to no singularity in their transfer functions.

Furthermore, the values of various transfer functions in the two kinds of solutions are listed in Tables~\ref{tab:TF123a} and~\ref{tab:TF123b} for the two selected cases, which is helpful to better comparison and analysis. As shown in Tables~\ref{tab:TF123a} and~\ref{tab:TF123b}, it is obviously seen that the difference between the two theoretical solutions becomes larger as the $\alpha$ value becomes smaller.  The values of transfer functions for quasi-linear terms are listed separately in Tables~\ref{tab:TF31a} and~\ref{tab:TF31b}. These values are different in the two theories except for the transfer functions $(\mathcal{W},W)$ in the third-order dispersion relation.

\begin{figure}
	\centering
	\subfigure[Case:~$\kappa h=\infty$]{
		\includegraphics[width=\textwidth]{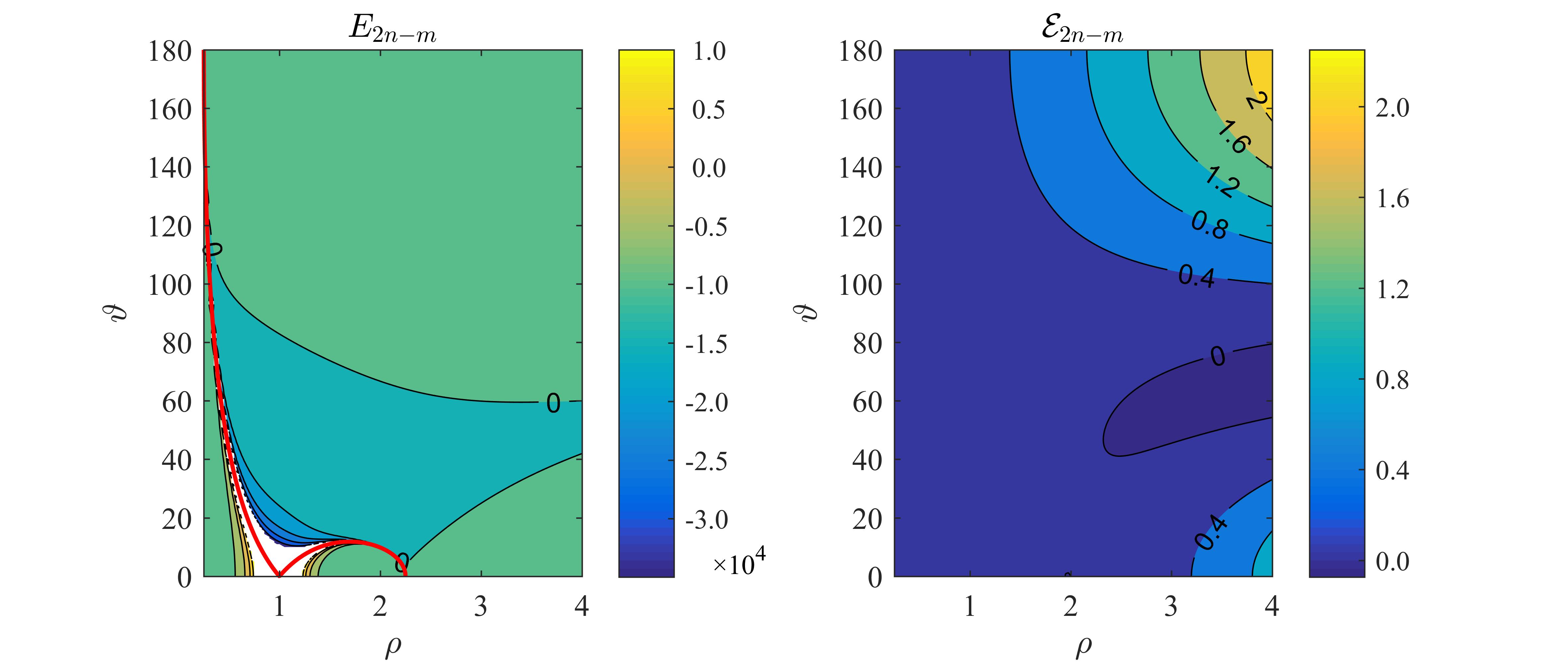} 
	}
	\quad
	\subfigure[Case:~$\kappa h=1$]{
		\includegraphics[width=\textwidth]{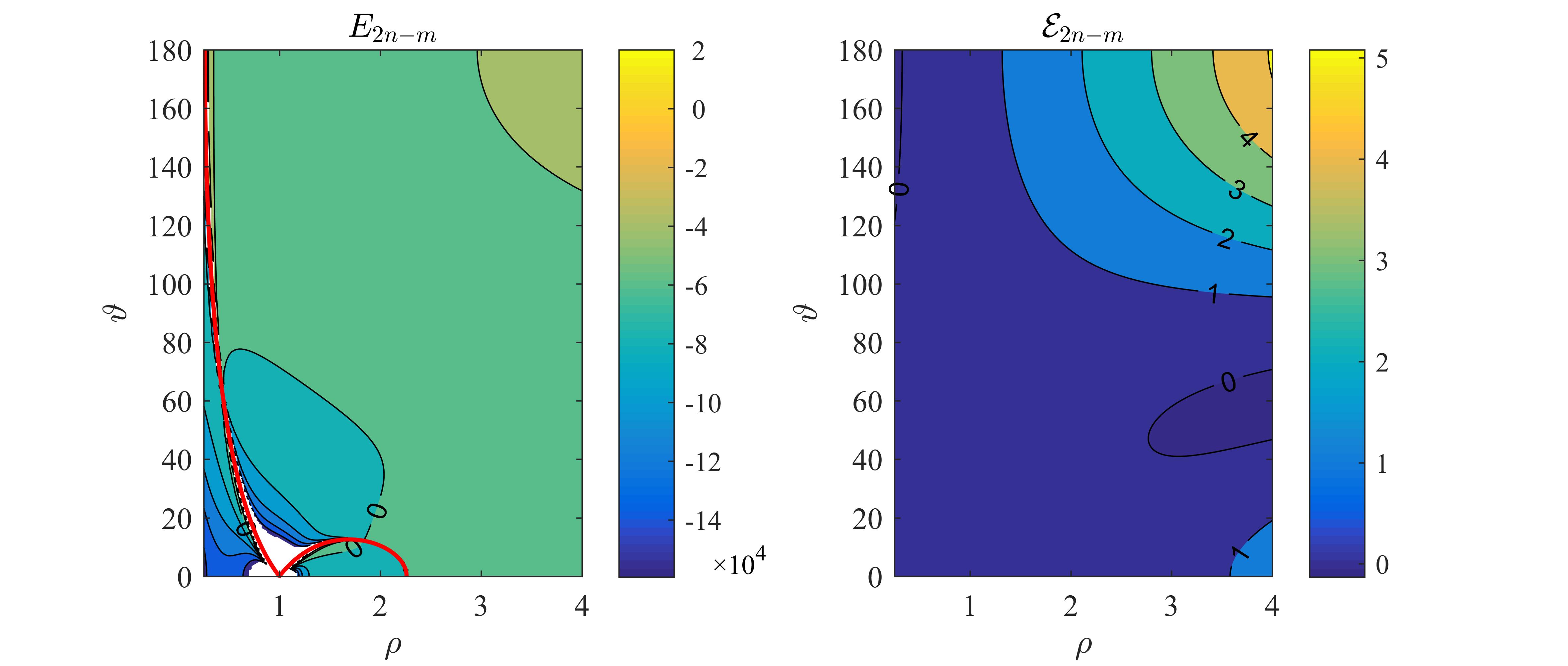}
	}
	\caption{Contours of dimensionless transfer functions $(E_{2n-m}, \mathcal{E}_{2n-m})$: $\mathcal{E}_{2n-m}$ corresponds to the Hamiltonian solution while ${E}_{2n-m}$ corresponds to the perturbation solution}
	\label{fig:E2nm}
\end{figure}

\begin{table}
	\begin{center}
		\begin{tabular}{lrrrrrrr}
			&\multicolumn{1}{c}{$\mathcal{E}$}&\multicolumn{1}{c}{${E}$}&                                  
			\multicolumn{1}{c}{$\mathcal{F}$}&\multicolumn{1}{c}{${F}$}&                           \multicolumn{1}{c}{$\mathcal{G}$}&\multicolumn{1}{c}{${G}$}&\multicolumn{1}{c}{$\alpha$}\\[3pt]
			${n+m}$  & 0.6061&  0.6061 &-0.4298& -0.4298 & 1.6050&  1.6050& \\
			${n-m}$  & 0.1156&  0.1156 & 0.1804&  0.1804 & 0.1322&  0.1322& \\
			${n+p}$  & 0.9229&  0.9229 & 0.0163&  0.0163 & 2.1032&  2.1032& \\
			${n-p}$  &-0.0108& -0.0108 & 0.8310&  0.8310 & 0.7308&  0.7308& \\
			${m+p}$  & 1.0443&  1.0443 & 0.1432&  0.1432 & 2.2782&  2.2782& \\
			${m-p}$  &-0.0276& -0.0276 & 0.8718&  0.8718 & 0.8198&  0.8198& \\
			${2n}$   & 1.1049&  1.1049 & 0.4392&  0.4392 & 2.4258&  2.4258& \\
			${2m}$   & 1.1248&  1.1248 & 0.3362&  0.3362 & 2.4191&  2.4191& \\
			${2p}$   & 1.1607&  1.1607 & 0.2479&  0.2479 & 2.4349&  2.4349& \\
			${n+m+p}$& 0.9662&  0.9662 &-0.2178& -0.2178 & 2.4489&  2.4489& ---\\
			${n+m-p}$& 0.2152&  1.8347 & 0.4678&  5.4678 &-0.6611&  5.1967& 12.83\\
			${n-m+p}$&       & -3.2378 &       & -7.0428 &       & -7.2807& 11.91\\
			${n-m-p}$&-0.6422& -3.3610 & 2.0830&  6.7635 & 1.2238&  7.0191& 13.03\\
			${n+2m}$ & 0.7497&  0.7497 &-0.2560& -0.2560 & 2.0559&  2.0559& \\
			${n-2m}$ &-0.3014& -1.6768 & 1.8302&  3.6041 & 0.5536&  3.6895& 22.00\\
			${2n+m}$ & 0.7202&  0.7202 &-0.2885& -0.2885 & 2.0144&  2.0144& \\
			${2n-m}$ & 0.4359& -1.0387 & 0.6862& -2.5561 &-0.8423& -2.5688& 29.04\\
			${n+2p}$ & 1.3762&  1.3762 &-0.1888& -0.1888 & 3.1777&  3.1777& \\
			${n-2p}$ &-1.0370& -7.7119 & 2.9537& 15.1620 & 1.9880& 15.5343& 7.68\\
			${2n+p}$ & 1.2916&  1.2916 &-0.2274& -0.2274 & 3.0585&  3.0585& \\
			${2n-p}$ & 0.4156& -2.4918 & 0.9927& -6.1825 &-0.9473& -6.3121& 16.37\\
			${m+2p}$ & 1.6300&  1.6300 &-0.1662& -0.1662 & 3.6124&  3.6124& \\
			${m-2p}$ &-1.1705&-14.0590 & 3.7809& 29.2052 & 2.3961& 29.5808& 4.70\\
			${2m+p}$ & 1.5785&  1.5785 &-0.1796& -0.1796 & 3.5388&  3.5388& \\
			${2m-p}$ & 0.4892& -7.5787 & 1.2831&-17.8940 &-1.1220&-18.1282& 7.17\\
			${3n}$   & 1.6285&  1.6285 &-0.1790& -0.1790 & 3.6781&  3.6781& \\
			${3m}$   & 1.7332&  1.7332 &-0.1641& -0.1641 & 3.8180&  3.8180& \\
			${3p}$   & 1.8901&  1.8901 &-0.1420& -0.1420 & 4.0409&  4.0409& \\
		\end{tabular}
		\caption{Case $\varepsilon=0$: comparison between the transfer functions $(E,F,G)$ in the perturbation solution and those $(\mathcal{E},\mathcal{F},\mathcal{G})$ in the Hamiltonian solution. Note that the second and third order transfer functions of surface elevation have the dimension ${\rm m}^{-1}$ and ${\rm m}^{-2}$ respectively, and the second and third order transfer functions of velocity potential have the dimension ${\rm s}^{-1}$ and $({\rm m\ s})^{-1}$ respectively.}
		\label{tab:TF123a}
	\end{center}
\end{table}
\begin{table}
	\begin{center}
		\begin{tabular}{lrrrrrrrr}
			&\multicolumn{1}{c}{$\mathcal{P}$}&\multicolumn{1}{c}{${P}$}&                              
			\multicolumn{1}{c}{$\mathcal{Q}$}&\multicolumn{1}{c}{${Q}$}& \multicolumn{1}{c}{$\mathcal{R}$}&\multicolumn{1}{c}{${R}$}&                  \multicolumn{1}{c}{$\mathcal{W}$}&\multicolumn{1}{c}{${W}$}  \\[3pt]
			$nn$  & -0.6105 & 0.7344 & -3.3204 &  0.0000 & -3.8537& -0.5333 &  3.4163 & 3.4163\\
			$nm$  & -0.3763 & 0.2955 & -1.6587 &  0.0000 & -3.0234& -1.3646 &  1.8925 & 1.8925\\
			$np$  & -0.7859 & 0.9347 & -4.8342 &  0.0000 & -5.4899& -1.5626 &  3.0140 & 3.0140\\
			$mn$  & -0.3852 & 0.1960 & -1.3685 &  0.0000 & -2.9232& -1.5546 &  1.8925 & 1.8925\\
			$mm$  & -0.6326 & 0.7470 & -3.2488 &  0.0000 & -3.9524& -0.7036 &  3.8437 & 3.8437\\
			$mp$  & -1.0326 & 1.1946 & -5.2448 &  0.0000 & -6.9942& -1.7494 &  3.6726 & 3.6726\\
			$pn$  & -0.9215 & 0.7383 & -3.7226 &  0.0000 & -5.7285& -2.0059 &  3.0140 & 3.0140\\
			$pm$  & -1.1500 & 1.0949 & -5.0349 &  0.0000 & -7.0274& -1.9925 &  3.6726 & 3.6726\\
			$pp$  & -0.6736 & 0.7745 & -4.3966 &  0.0000 & -4.1416& -0.8938 &  4.4028 & 4.4028 
		\end{tabular}
		\caption{Case $\varepsilon=0$: comparison of the quasi-linear transfer functions, $(P,Q,R)$ for perturbation solution and $(\mathcal{P},\mathcal{Q},\mathcal{R})$ for Hamiltonian solution. Note that $\mathcal{P}/P$ and $\mathcal{W}/W$ have the dimension ${\rm m}^{-2}$ while $\mathcal{Q}$ and $\mathcal{R}/R$ have the dimension $({\rm m\;s})^{-1}$.} 
		\label{tab:TF31a}
	\end{center}
\end{table}
\begin{table} 
	\begin{center}
		\begin{tabular}{lrrrrrrr}
			&\multicolumn{1}{c}{$\mathcal{E}$}&\multicolumn{1}{c}{${E}$}&                                  
			\multicolumn{1}{c}{$\mathcal{F}$}&\multicolumn{1}{c}{${F}$}&                           \multicolumn{1}{c}{$\mathcal{G}$}&\multicolumn{1}{c}{${G}$}&\multicolumn{1}{c}{$\alpha$}\\[3pt]
			${n+m}$  & 0.6061&  0.6061 &-0.4298& -0.4298 & 1.6050&  1.6050& \\
			${n-m}$  & 0.1156&  0.1156 & 0.1804&  0.1804 & 0.1322&  0.1322& \\
			${n+p}$  & 1.1616&  1.1616 & 0.2909&  0.2909 & 2.4407&  2.4407& \\
			${n-p}$  &-0.3795& -0.3795 & 5.2489&  5.2489 & 5.0857&  5.0857& \\
			${m+p}$  & 0.7163&  0.7163 &-0.4358& -0.4358 & 1.7622&  1.7622& \\
			${m-p}$  & 0.1014&  0.1014 & 0.4952&  0.4952 & 0.3801&  0.3801& \\
			${2n}$   & 1.1049&  1.1049 & 0.4392&  0.4392 & 2.4258&  2.4258& \\
			${2m}$   & 1.1248&  1.1248 & 0.3362&  0.3362 & 2.4191&  2.4191& \\
			${2p}$   & 1.2229&  1.2229 & 0.1673&  0.1673 & 2.4804&  2.4804& \\
			${n+m+p}$& 0.8795&  0.8795 &-0.2231& -0.2231 & 2.2927&  2.2927& ---\\
			${n+m-p}$& 0.1276&110.0537 & 0.3720&307.7687 &-0.3567&307.4315& 0.07\\
			${n-m+p}$&       & -1.9810 &       & -4.1171 &       & -4.2525& 17.17\\
			${n-m-p}$&-1.4471& 17.9716 & 2.1020&-37.3346 & 2.9887&-36.6121& 1.19\\
			${n+2m}$ & 0.7497&  0.7497 &-0.2560& -0.2560 & 2.0559&  2.0559& \\
			${n-2m}$ &-0.3014& -1.6768 & 1.8302&  3.6041 & 0.5536&  3.6895& 17.03\\
			${2n+m}$ & 0.7202&  0.7202 &-0.2885& -0.2885 & 2.0144&  2.0144& \\
			${2n-m}$ & 0.4359& -1.0387 & 0.6862& -2.5561 &-0.8423& -2.5688& 22.48\\
			${n+2p}$ & 1.9634&  1.9634 &-0.1418& -0.1418 & 4.1497&  4.1497& \\
			${n-2p}$ &-3.5957& 21.3470 & 6.0276&-40.5884 & 6.8090&-39.4276& 1.71\\
			${2n+p}$ & 1.7908&  1.7908 &-0.1650& -0.1650 & 3.9051&  3.9051& \\
			${2n-p}$ & 0.3366&  4.4939 & 1.0646& 13.6571 &-1.0395& 13.1870& 2.51\\
			${m+2p}$ & 1.0510&  1.0510 &-0.1625& -0.1625 & 2.5863&  2.5863& \\
			${m-2p}$ &-0.6406& -3.9095 & 2.0194&  7.1201 & 1.0861&  7.3913& 12.13\\
			${2m+p}$ & 0.9429&  0.9429 &-0.2186& -0.2186 & 2.4201&  2.4201& \\
			${2m-p}$ & 0.4475& -1.2636 & 0.7060& -2.9755 &-0.8630& -3.0122& 23.96\\
			${3n}$   & 1.6285&  1.6285 &-0.1790& -0.1790 & 3.6781&  3.6781& \\
			${3m}$   & 1.7332&  1.7332 &-0.1641& -0.1641 & 3.8180&  3.8180& \\
			${3p}$   & 2.1455&  2.1455 &-0.1126& -0.1126 & 4.4103&  4.4103& \\
		\end{tabular}
		\caption{Case $\varepsilon=0.95$: comparison between the transfer functions $(E,F,G)$ in Perturbation solutions and those $(\mathcal{E},\mathcal{F},\mathcal{G})$ in Hamiltonian solutions. Note that the second and third order transfer functions of surface elevation have the dimension ${\rm m}^{-1}$ and ${\rm m}^{-2}$ respectively, and the second and third order transfer functions of velocity potential have the dimension ${\rm s}^{-1}$ and $({\rm m\;s})^{-1}$ respectively.}
		\label{tab:TF123b}
	\end{center}
\end{table}
\begin{table}
	\begin{center}
		\begin{tabular}{lrrrrrrrr}
			&\multicolumn{1}{c}{$\mathcal{P}$}&\multicolumn{1}{c}{${P}$}&                              
			\multicolumn{1}{c}{$\mathcal{Q}$}&\multicolumn{1}{c}{${Q}$}& \multicolumn{1}{c}{$\mathcal{R}$}&\multicolumn{1}{c}{${R}$}&                  \multicolumn{1}{c}{$\mathcal{W}$}&\multicolumn{1}{c}{${W}$}  \\[3pt]
			$nn$  & -0.6105 & 0.7344 & -3.3204 &  0.0000 & -3.8537& -0.5333 &  3.4163 & 3.4163\\
			$nm$  & -0.3763 & 0.2955 & -1.6587 &  0.0000 & -3.0234& -1.3646 &  1.8925 & 1.8925\\
			$np$  &  0.3760 & 0.8933 & -3.2886 &  0.0000 & -3.8167& -2.7603 &  3.0041 & 3.0041\\
			$mn$  & -0.3852 & 0.1960 & -1.3685 &  0.0000 & -2.9232& -1.5546 &  1.8925 & 1.8925\\
			$mm$  & -0.6326 & 0.7470 & -3.2448 &  0.0000 & -3.9524& -0.7036 &  3.8437 & 3.8437\\
			$mp$  & -0.5156 & 0.4464 & -2.2654 &  0.0000 & -3.9886& -1.7232 &  2.6511 & 2.6511\\
			$pn$  & -3.2591 & 0.7075 & -8.4115 &  0.0000 &-12.1494& -3.7379 &  3.0041 & 3.0041\\
			$pm$  & -0.5310 & 0.1614 & -1.4684 &  0.0000 & -3.7685& -2.3001 &  2.6511 & 2.6511\\
			$pp$  & -0.7477 & 0.8304 & -4.6210 &  0.0000 & -4.4808& -1.1342 &  5.2305 & 5.2305 
		\end{tabular}
		\caption{Case $\varepsilon=0.95$: comparison of the quasi-linear transfer functions, $(P,Q,R)$ for Perturbation solutions and $(\mathcal{P},\mathcal{Q},\mathcal{R})$ for Hamiltonian solutions. Note that $\mathcal{P}/P$ and $\mathcal{W}/W$ have the dimension ${\rm m}^{-2}$ while $\mathcal{Q}$ and $\mathcal{R}/R$ have the dimension $({\rm m\;s})^{-1}$.}
		\label{tab:TF31b}
	\end{center}
\end{table}
\begin{table}
	\begin{center}
		\begin{tabular}{cccccc}
			& $\theta_{n+ m}$ & $\theta_{n-m}$& $\theta_{n+m+p}$&$\theta_{n+m-p}$&$\theta_{n-m-p}$\\[3pt]
	 $\mathcal{E}/E$&$ = $& $=$ & $=$ &$\neq$ & $\neq$ \\ 
	 $\mathcal{F}/F$&$ = $& $=$ & $=$ &$\neq$ & $\neq$ \\ 
	 $\mathcal{G}/G$&$ = $& $=$ & $=$ &$\neq$ & $\neq$ \\ 
	 \hline
	 $\mathcal{W}/W$&\multicolumn{5}{c}{$\mathcal{W}_{nm}=W_{nm}$}          
	    \end{tabular}
		\caption{Comparison of transfer functions in two solutions: ($\mathcal{E},\mathcal{F},\mathcal{G}$) correspond to the Hamiltonian solution while (${E},{F},{G}$) correspond to the perturbation solution} 
	    \label{tab:comparison}
	\end{center}
\end{table}
\subsection{Comparison of transfer functions} 

Through the above simple trichromatic example, it is found that some of the transfer functions in the two solutions give the same value, although the analytical forms of these transfer functions are different. Intuitively, this does not seem to be an accidental phenomenon, but an inevitable result. In other words, these transfer functions having different forms are equivalent. To confirm this conjecture, we use Maxima (an open-source computer algebra system) to compare the transfer functions in the two theories one by one. The result of comparison is presented in Table~\ref{tab:comparison}. It shows the first- and second-order transfer functions are completely equal, and in the third-order part only the transfer function of $\theta_{n+m+p}$ terms are equal. In addition, the third-order dispersion relations given by the two solutions are also equivalent. 

For those unequal transfer functions, we shall focus on a simple case involving only two wavevector in which the wavevector ${\bf k}_n$ is computed twice, defined by
\begin{equation*}
	{\bf k}_n=\kappa(1,0),\quad {\bf k}_m=\kappa(\rho\cos\vartheta,\rho\sin\vartheta),
\end{equation*}
where $\vartheta$ is the angle between the two wavevector and $\rho=k_m/k_n$. The contours of dimensionless transfer functions of surface elevation ($E_{2n-m}$, $\mathcal{E}_{2n-m}$ normalized by $\kappa^2$) are shown in Figure~\ref{fig:E2nm}. The upper diagrams of this figure correspond to the case of deep water $\kappa h=\infty$ and the lower diagrams correspond to the case of finite depth $\kappa h=1$. As can be seen from the figure, there is a collapse that take place near the red solid line for $E_{2n-m}$, resulting in some unreasonable values in the white area. The red solid line represents a resonance curve which is identical to the ‘figure-eight’ curve as shown by \citet{phillips1960dynamics}. In contrast, $\mathcal{E}_{2n-m}$ in the Hamiltonian solution is not affected by resonance singularity. Furthermore, the contour of this transfer function in the case of finite depth is similar to that in the deep-water case, but with a expansion of the range. It indicates that a decrease of relative water depth leads to an increase of wave nonlinearity.

\section{Some consequences}\label{sec:consequences}
\subsection{The non-uniqueness of induce mean flow and mean surface}\label{subsec:non-uniqueness}
As mentioned in subsection \ref{subsec:2Psolution}, the induced mean flow is compensated by ${\bf U}$ to meet the Stokes first definition, and the mean surface  is counterbalanced by the coefficient $C_1$. Note that the induced mean flow and mean surface do not appear in our second-order solution, because we chose the MWL as the datum and the Stokes' first wave velocity definition. If we choose the mean energy level (MEL) as the datum, the mean surface will appear, which is also called set-down. The phenomenon that different datums lead to different solutions, for the third-order Stokes wave, is discussed in detail by \cite{Song2012OnTU}.

There exist two limits ($E_{n-n}$ and $F_{n-n}\theta_{n-n}$) that make the non-uniqueness of induced mean flow and mean surface. 
Defining
\begin{equation*}
	\cos\theta_{n21}=\lim_{\bm{\varepsilon}_1,\bm{\varepsilon}_2\rightarrow 0}\frac{{\bf k}_n\cdot({\bm\varepsilon}_2-{\bm\varepsilon}_1)}{k_n|\bm\varepsilon_2-\bm{\varepsilon}_1|},
\end{equation*}
and using
\begin{equation*}
	\lim_{\bm{\varepsilon}_1,\bm{\varepsilon}_2\rightarrow 0}\left[  \omega({\bf k}_n+\bm{\varepsilon}_1)- \omega({\bf k}_n+\bm{\varepsilon}_2)\right]  =\bm{e}_{k_n}\cdot(\bm{\varepsilon}_1-\bm{\varepsilon}_2)\;{c}_{gn}
\end{equation*}
where $\bm{\varepsilon}_1, \bm{\varepsilon}_2$ are perturbed vectors which are allowed to approach zero in any direction, $\mathbf{e}_{k_n}={\bf k}_n/k_n$ is unit vector in the wave propagation direction, ${c}_{gn}$ is group velocity and related to phase velocity ${c}_{pn}$:
\[
{c}_{gn}= \frac{1}{2}{c}_{pn}\left[  1+\kappa_{n}\left(\frac{1}{\sigma_n}-\sigma_n\right)\right],\quad \sigma_n=\tanh(k_nh),
\]
and $\kappa_{n}=k_nh$ is a relative depth. Then, using the second-order part of perturbation solution and some mathematical operations, the two limits become
\begin{subequations}
	\begin{equation}
		{E}_{n- n}=-\frac{k_n}{4}\frac{gh}{gh-c_{gn}^2\cos^2\theta_{n21}}\left[  2\frac{c_{gn}}{\kappa_n c_{pn}}\cos^2\theta_{n21}+\left(\frac{1}{\sigma_n}-\sigma_n\right)\right] 
	\end{equation}
    \begin{equation}
    	F_{n-n}\theta_{n-n}={\bf U}_n\cdot{\bf x}+C_{2n}\;t 
    \end{equation}
\end{subequations}
where
\[
{\bf U}_{n}=-\frac{{\omega_n}}{4}\frac{g}{gh-c_{gn}^2\cos^2\theta_{n21}}\left[\frac{c_{gn}}{c_{n}}\left( \frac{1}{\sigma_n}-\sigma_n\right) +\frac{2}{\sigma_n} \right]\cos\theta_{n21}\lim_{\bm{\varepsilon}_1,\bm{\varepsilon}_2\rightarrow 0}\frac{\bm{\varepsilon}_2-\bm{\varepsilon}_1}{|\bm{\varepsilon}_2-\bm{\varepsilon}_1|},
\]
\[
C_{2n}=\frac{k_n}{4}\frac{g}{gh-c_{gn}^2\cos^2\theta_{n21}}\left[c_{gn}^2\left( \frac{1}{\sigma_n}-\sigma_n\right)+\frac{2gc_{gn}}{k_nc_{pn}} \right]\cos^2\theta_{n21},
\]
It can be seen from the above analysis that the induce mean flow and the mean surface have non-unique values and depend on the direction of approaching the limit $\cos\theta_{n21}$. Furthermore, if the coordinate system is fixed at the MEL, these limits will also appear in quasi-linear terms, including the nonlinear dispersion relation. The non-uniqueness in nonlinear dispersion relations was discussed by \cite{stiassnie2009zakharov} in the context of the Zakharov equation. 

\subsection{Conservation properties}
For the Hamiltonian theory of water waves, there are some conserved quantities. For instance, the total energy and the horizontal momentum. For more details on conserved quantities, we refer to \cite{benjamin1982hamiltonian}. In addition, the four-wave system conserves the wave action, i.e.,
\begin{equation}\label{eq:conserv-waveaction}
	\frac{\partial J}{\partial t}=0,\qquad J=\int \omega({\bf k})n({\bf k})d{\bf k}
\end{equation}
where the wave action spectrum $n({\bf k})$ is defined by $\overline{ {A({\bf k})A^*({\bf k}_1)}}=n({\bf k})\delta({\bf k}-{\bf k}_1)$. By using 
\[
E({\bf k})=\frac{\omega({\bf k})}{g}n({\bf k}),
\]
the conserved quantities $J$ corresponds to the second moment of sea surface or the zero moment of wavenumber spectrum $E({\bf k})$, namely
 $$\overline{\zeta^2}=\int E({\bf k})d{\bf k}.$$
Note that the wavenumber spectrum $E({\bf k})$ is the sum of the first $S({\bf k})$ and second order spectrum. The derivation of the second-order spectrum can be referred to \cite{janssen2009some}. According to the conserved law \eqref{eq:conserv-waveaction}, $\overline{\zeta^2}$ is independent of time and equal to the second moment of initial sea surface, which is assumed to be a Gaussian random process. Thus, there is
\begin{equation}\label{eq:conserv-P.E}
	\overline{\zeta^2}=\overline{\zeta_1^2}=m_0,
\end{equation}
where $m_0=\int S({\bf k})d{\bf k}$ is the zero moment of wave spectrum.
This indicate that the contribution of nonlinear parts does not change the mean potential energy. Similarly, as for the mean kinetic energy there exists
\begin{equation}\label{eq:conserv-K.E}
	\overline{|\psi\zeta_t|}=\overline{|\psi_1\zeta_{1t}|}=m_0.
\end{equation}
Then, substituting the third-order Hamiltonian solution into $\overline{\zeta^2}$ and $\overline{|\psi\zeta_t|}$ up to $\mathcal{O}(\epsilon^4)$, gives
\begin{subequations}\label{eq:conserv-P.EandK.E}
	\begin{equation}
		\overline{\zeta^2}=\int S({\bf k})d{\bf k}+\int \mathcal{T}_{12}^{(1)} S({\bf k}_1)S({\bf k}_2)d{\bf k}_{12},
	\end{equation}
	\begin{equation}
		\overline{|\psi\zeta_t|}=\int S({\bf k})d{\bf k}+\int \mathcal{T}_{12}^{(2)} S({\bf k}_1)S({\bf k}_2)d{\bf k}_{12},
	\end{equation}
\end{subequations}
where 
\begin{align*}
	\mathcal{T}_{12}^{(1)}&=4\left( \mathcal{E}_{n+m}^2+\mathcal{E}_{n-m}^2+\mathcal{P}_{nm}\right), \\
	\mathcal{T}_{12}^{(2)}&=4\omega_{n+m}\mathcal{E}_{n+m}\mathcal{G}_{n+m}+4\omega_{n-m}\mathcal{E}_{n-m}\mathcal{G}_{n-m}+2g\mathcal{P}_{nm}+2\omega_{n}\mathcal{R}_{nm}.
\end{align*}
In order to satisfy \eqref{eq:conserv-P.E} and \eqref{eq:conserv-K.E}, that is, the second integral must be zero, one finds that $\mathcal{T}_{12}^{()}$  is antisymmetric, i.e., $\mathcal{T}_{12}^{()}=-\mathcal{T}_{21}^{()}$ and $\mathcal{T}_{11}^{()}=0$.
Note that there is more than two such transfer functions, and different conserved quantities lead to different antisymmetric transfer functions $\mathcal{T}_{12}^{()}$.

\subsection{Third-order Stokes wave}
In one sense, the solutions presented in this paper are generalized third-order Stokes wave. Considering an unidirectional case with a single frequency, the solutions reduce to a third-order Stokes wave. Base on the Hamiltonian solutions, we obtain
\begin{subequations}
	\begin{align}
		k\zeta&=(1+\alpha_1\epsilon^2)\epsilon\cos\theta+\beta_1 \epsilon^2\cos2\theta+\gamma_1 \epsilon^3\cos3\theta\\
		\frac{k^2}{\omega}\psi&=\left( {1}/{\sigma}+\alpha_2\epsilon^2\right) \epsilon\sin\theta+\beta_2\epsilon^2\sin2\theta+\gamma_2\epsilon^3\sin3\theta\\
		\frac{k^2}{\omega}\phi&=(1/\sigma+\alpha_3\epsilon^2)\epsilon f(Z)\sin\theta+\beta_3\epsilon^2f(2Z)\sin 2\theta+ \gamma_3 a^3f(3Z)\sin3\theta
	\end{align}
\end{subequations}
where $f(nZ)=\cosh{nk(z+h)}/\cosh nkh$, $\sigma=\tanh kh$, $\epsilon=ka$ and
\[
\alpha_1=-\frac{1}{2}\beta^2,\quad \beta_1=\frac{3-\sigma^2}{4\sigma^3},\quad \gamma_1=\frac{3}{64}\frac{8+(1-\sigma^2)^3}{\sigma^6}
\]
\[
\alpha_2=\frac{(\sigma^2-3)(2\sigma^4+\sigma^2+3)}{32\sigma^7}, \quad \beta_2=\frac{3+\sigma^4}{8\sigma^4},\quad\gamma_2=\gamma_3+\frac{9-6\sigma^2}{8\sigma^3} 
\]
\[
\alpha_3=\alpha_2-\frac{3-4\sigma^2}{8\sigma^3} ,\quad \beta_3=\frac{3(1-\sigma^4)}{8\sigma^4},\quad \gamma_3=\frac{(\sigma^2-1)(13\sigma^2-9)(1+3\sigma^2)}{64\sigma^7}
\]
For deep water $\kappa=\infty$, the expression of the third-order Stokes wave become
\begin{subequations}
	\begin{align}
		k\zeta&=(1-\frac{1}{8}\epsilon^2)\epsilon\cos\theta+\frac{1}{2} \epsilon^2\cos2\theta+\frac{3}{8} \epsilon^3\cos3\theta\\
		\frac{k^2}{\omega}\psi&=(1-\frac{3}{8}\epsilon^2)\epsilon\sin\theta+\frac{1}{2} \epsilon^2\sin2\theta+\frac{3}{8} \epsilon^3\sin3\theta\\
		\frac{k^2}{\omega}\phi&=(1-\frac{1}{4}\epsilon^2)\epsilon\; e^{kz}\sin\theta
	\end{align}
\end{subequations}

\subsection{Volume flux}\label{subsec:volume flux}
The time-averaged volume flux vector is defined by
\begin{equation}
	{\bf M}=\overline{\int_{-h}^{\zeta}\nabla_{\bf x}\phi\; dz}=\int_{-h}^0\overline{\nabla_{\bf x}\phi}\;dz+\overline{\int_{0}^{\zeta}\nabla_{\bf x}\phi\; dz}
\end{equation}
The first integral is zero and the second integral can be evaluated by using Taylor series expansions. Accurate to $\mathcal{O}(\epsilon^4)$, we obtain
\begin{equation}
	{\bf M}={\bf U}+\sum_{n}\frac{g{\bf k}_n}{2\omega_n}a_n^2+\sum_{n,m}\mathcal{V}_{nm}a_n^2a_m^2 +\mathcal{O}(\epsilon^6),
\end{equation}
where
\begin{align}
\mathcal{V}_{nm}=& \mathcal{E}_{n+m}\mathcal{F}_{n+m}({\bf k}_n+{\bf k}_m)+\mathcal{E}_{n-m}\mathcal{F}_{n-m}({\bf k}_n-{\bf k}_m)+\frac{1}{2}{\bf k}_n\mathcal{Q}_{nm}+\frac{1}{2}g\frac{{\bf k}_n}{\omega_n}\mathcal{P}_{nm}\notag\\
&+\frac{1}{4}\left( \mathcal{F}_{n+m}({\bf k}_n+{\bf k}_m)|{\bf k}_n+{\bf k}_m|+\mathcal{F}_{n-m}({\bf k}_n-{\bf k}_m)|{\bf k}_n-{\bf k}_m|\right) \notag\\
&+\frac{1}{2}{\omega_m}{\bf k}_n\left( \mathcal{E}_{n+m}+\mathcal{E}_{n-m}\right)+\frac{1}{8}\frac{gk_n^2}{\omega_n}{\bf k}_n 
\end{align}
In the case of closed wave tanks, the time-averaged volume flux must be zero, resulting in a wave-induced return current. This corresponds to Stokes' second wave velocity definition and this return current can be determined by
\begin{equation}
{\bf U}=-g\int\frac{{\bf k}}{\omega({\bf k})}S({\bf k})d{\bf k}-4\int \mathcal{V}_{nm}({\bf k}_1,{\bf k}_2)S({\bf k}_1)S({\bf k}_2)d{\bf k}_{12}
\end{equation}
\subsection{Skewness and excess kurtosis}
Skewness is a measure of vertical asymmetry of the sea surface, which is important in the determination of the sea state bias as experienced by a radar altimeter, and the excess kurtosis measures the probability of extreme waves. They are defined as follows
\begin{equation}\label{eq:SkewKurt}
	\lambda_3=\frac{\overline{\zeta^3}}{\left( \overline{\zeta^2}\right)^{3/2}},\qquad \lambda_4=\frac{\overline{\zeta^4}}{\left( \overline{\zeta^2}\right)^2 }-3.
\end{equation}
By substituting the Hamiltonian solution \eqref{eq:surface} into \eqref{eq:SkewKurt} and using $\overline{\zeta^2}=m_0$, we obtain
\begin{align}
	\lambda_3=&\frac{6}{m_0^{3/2}}\int\left( \mathcal{E}_{1+2}+\mathcal{E}_{1-2}\right)S({\bf k}_1)S({\bf k}_2) d{\bf k}_{12}\notag\\
	&+\frac{1}{m_0^{3/2}}\int  \left\lbrace  72\mathcal{E}_{2+3}\left( \mathcal{E}_{1+2+3}+\mathcal{E}_{1+2-3}+\mathcal{E}_{1-2-3}\right)\right. \notag\\
	&+16\mathcal{E}_{2-3}\left( \mathcal{E}_{1+2-3}+\mathcal{E}_{1-2-3}\right)\notag\\
	&\left. +48 \mathcal{E}_{1+2}\mathcal{E}_{2+3}\mathcal{E}_{1-3}-16\mathcal{E}_{1-2}\mathcal{E}_{2-3}\mathcal{E}_{1-3}\right\rbrace  S({\bf k}_1)S({\bf k}_2)S({\bf k}_3)d{\bf k}_{123}
\end{align}
\begin{align}
	\lambda_4=&\frac{24}{m_0^2}\int \left\lbrace 2\left( \mathcal{E}_{2+3}+ \mathcal{E}_{2-3}\right)\left(\mathcal{E}_{1+3}+ \mathcal{E}_{1-3} \right)\right. \notag\\
	& \left. +\left( \mathcal{E}_{1+2+3}+\mathcal{E}_{1+2-3}+\mathcal{E}_{1-2-3}\right)\right\rbrace   S({\bf k}_1)S({\bf k}_2)S({\bf k}_3)d{\bf k}_{123}
\end{align}
This is consistent with the expression of skewness and excess kurtosis in \cite{janssen2009some}, except that we give the contribution of the third-order part of skewness. For the case of deep water with a single frequency, the skewness and excess kurtosis become
\[
\lambda_3=3\epsilon+9\epsilon^3,\qquad \lambda_4=18\epsilon^2,
\]
where $\epsilon=km_0^{1/2}$.

\section{Conclusions} \label{sec:conclusion}
This study derived a third-order steady-state solution of surface gravity waves, namely the approximate analytical solution of irrotational Euler equations. Explicit expressions for free-surface velocity potential, surface elevation and the vertical variation of velocity potential were provided in this analytical solution, as well as the third-order dispersion relation. Two approaches were used to obtain the approximate solution: one is the perturbation expansion technique based on the original water wave equations, the other is the canonical transformation in the context of the Hamiltonian theory of water wave. A comprehensive comparison of two kinds of analytical solutions (i.e., perturbation solution and Hamiltonian solution) was carried out. It is found that the first- and second-order parts in the two solutions are absolutely equivalent, including the $\theta_{n+m+p}$ term in the third-order part. The nonlinear dispersion relation is also the same in the two types of solutions. 

Compared with the perturbation solution, the Hamiltonian solution has two main advantages. One is that the Hamiltonian solution removes the barriers that the perturbation solution will break down due to inherent singularities in the transfer functions. The other is the Hamiltonian solution inherits important properties of the original water wave system. For example, the mean potential energy is equal to that in the initial state which is assumed to follow the linear random wave theory. Essentially, this is an embodiment of conserved quantities in the Hamiltonian form. All findings reported here show that the Hamiltonian solution is a more reasonable analytical theory to describe steady-state random wave field. This Hamiltonian analytical solution could not only quickly simulate a nonlinear random wave field based on FFT, but also serve as a theoretical basis for research topics such as wave statistical distributions. Finally, based on the Hamiltonian solution, the explicit expressions for some statistics were given, such as the volume flux, skewness, and excess kurtosis. 
\appendix
\section{Transfer coefficients in the perturbation solution}\label{appA}
\subsection{Second order}
The second-order super-harmonic  transfer coefficients in \eqref{eq:2Psol} are
\begin{subequations}\label{eq:secondTranfunsup}
	\begin{align}
	\mathcal{A}_{n+m}=&-\frac{1}{4\omega_{n}\omega_{m}}\Big(g^2\omega_{n}(2{\bf{k}}_n \cdot{\bf{k}}_m+k_m^2)+ 
	g^2\omega_{m}(2{\bf{k}}_n \cdot{\bf{k}}_m+k_n^2)\nonumber \\
	&-\omega_{n}\omega_{m}\omega_{n+m}(\omega_{n+ m}^2-\omega_{n}\omega_{m})\Big),\\
	\mathcal{B}_{n+m}=&-\frac{g^2{\bf{k}}_n\cdot{\bf{k}}_m}{4\omega_n\omega_m}+\frac{1}{4}(\omega_n^2+\omega_m^2+\omega_{n}\omega_{m}), \\
	\mathcal{C}_{n+ m} =&\ \ \frac{\omega_{n+ m}}{4}.
	\end{align}
\end{subequations}
Note that the second-order transfer coefficients are functions of the following arguments
\begin{subequations}\label{eq:secondTranfun}
	\begin{alignat}{1}
	A_{n\pm m}&=\varLambda_2 \left\lbrace\left[ \omega_{n},{\bf{k}}_n,k_n\right];\left[  \pm\omega_{ m},\pm{\bf{k}}_m,k_m\right] ;\left[ \omega_{n\pm m} \right]\right\rbrace      \label{subeq:Etransfun}\\
	B_{n\pm m}&=\varGamma_2\left\lbrace\left[ \omega_{n},{\bf{k}}_n,k_n\right];\left[  \pm\omega_{ m},\pm{\bf{k}}_m,k_m\right] \right\rbrace \\
	C_{n\pm m}&=\varXi_2\left\lbrace\left[ \omega_{n\pm m} \right]\right\rbrace  
	\end{alignat}
\end{subequations}
So the second-order sub-harmonic  transfer coefficients can be determined by switching the arguments in (\ref{eq:secondTranfunsup}).

\subsection{Third order}
The third-order transfer coefficients for terms $\theta_{n+m+p}$ are as follow:
\begin{subequations}\label{eq:ABCnmp}
	\begin{align}
	 \mathcal{A}_{n+m+p}=&-\frac{g}{24}\bigg( \omega_{n}({\bf k}_n\cdot{\bf k}_m+{\bf k}_n\cdot{\bf k}_p+k_n^2)\notag\\
	&+\omega_{m}({\bf k}_m\cdot{\bf k}_n+{\bf k}_m\cdot{\bf k}_p+k_m^2)\notag\\
	&+\omega_{p}({\bf k}_p\cdot{\bf k}_n+{\bf k}_p\cdot{\bf k}_m+k_p^2)\notag\\
	&+\frac{\omega_{n+m+p}}{\omega_{n}}(\omega_{m}{\bf k}_m\cdot{\bf k}_n+\omega_{p}{\bf k}_p\cdot{\bf k}_n-\omega_{n+m+p}k_n^2)\notag\\
	&+\frac{\omega_{n+m+p}}{\omega_{m}}(\omega_{n}{\bf k}_n\cdot{\bf k}_m+\omega_{p}{\bf k}_p\cdot{\bf k}_m-\omega_{n+m+p}k_m^2)\notag\\
	&+\frac{\omega_{n+m+p}}{\omega_{p}}(\omega_{n}{\bf k}_n\cdot{\bf k}_p+\omega_{m}{\bf k}_m\cdot{\bf k}_p-\omega_{n+m+p}k_p^2)\bigg)\notag\\
	&+\frac{E_{n+m}}{6}\bigg(\omega_{p}^2\omega_{n+m+p}-\frac{g^2}{\omega_{p}}({\bf k}_n\cdot{\bf k}_p+{\bf k}_m\cdot{\bf k}_p+k_p^2)\bigg)\notag\\
	&+\frac{E_{n+p}}{6}\bigg(\omega_{m}^2\omega_{n+m+p}-\frac{g^2}{\omega_{m}}({\bf k}_n\cdot{\bf k}_m+{\bf k}_p\cdot{\bf k}_m+k_m^2)\bigg)\notag\\
	&+\frac{E_{m+p}}{6}\bigg(\omega_{n}^2\omega_{n+m+p}-\frac{g^2}{\omega_{n}}({\bf k}_m\cdot{\bf k}_n+{\bf k}_p\cdot{\bf k}_n+k_n^2)\bigg)\notag\\
	&-\frac{F_{n+m}}{6}\bigg(g\big({\bf k}_n\cdot{\bf k}_p+{\bf k}_m\cdot{\bf k}_p+k_{n+m}^2\notag\\
	&+\frac{\omega_{n+m+p}}{\omega_{p}}({\bf k}_n\cdot{\bf k}_p+{\bf k}_m\cdot{\bf k}_p)\big)-k_{n+m}\tanh(k_{n+m}h)\omega_{n+m+p}^2\bigg)\notag\\
	&-\frac{F_{n+p}}{6}\bigg(g\big({\bf k}_n\cdot{\bf k}_m+{\bf k}_p\cdot{\bf k}_m+k_{n+p}^2\notag\\
	&+\frac{\omega_{n+m+p}}{\omega_{m}}({\bf k}_n\cdot{\bf k}_m+{\bf k}_p\cdot{\bf k}_m)\big)-k_{n+p}\tanh(k_{n+p}h)\omega_{n+m+p}^2\bigg)\notag\\
	&-\frac{F_{m+p}}{6}\bigg(g\big({\bf k}_m\cdot{\bf k}_n+{\bf k}_p\cdot{\bf k}_n+k_{m+p}^2\notag\\
	&+\frac{\omega_{n+m+p}}{\omega_{n}}({\bf k}_m\cdot{\bf k}_n+{\bf k}_p\cdot{\bf k}_n)\big)-k_{m+p}\tanh(k_{m+p}h)\omega_{n+m+p}^2\bigg)\label{subeq:Anmp}\\
	 \mathcal{B}_{n+m+p}=&-\frac{g}{24}\bigg(\frac{1}{\omega_{m}}(\omega_{p}{\bf k}_m\cdot{\bf k}_p+\omega_{n}{\bf k}_m\cdot{\bf k}_n-\omega_{n+m+p}k_m^2)\notag\\
	&+\frac{1}{\omega_{p}}(\omega_{m}{\bf k}_m\cdot{\bf k}_p+\omega_{n}{\bf k}_n\cdot{\bf k}_p-\omega_{n+m+p}k_p^2)\notag\\
	&+\frac{1}{\omega_{n}}(\omega_{p}{\bf k}_p\cdot{\bf k}_n+\omega_{m}{\bf k}_m\cdot{\bf k}_n-\omega_{n+m+p}k_n^2)\bigg)\notag\\ 
	&+\frac{1}{6}\left( E_{n+m}\omega_{p}^2+E_{n+p}\omega_{m}^2+E_{m+p}\omega_{n}^2\right)\notag\\
	&+\frac{1}{6}F_{n+m}\left( k_{n+m}\tanh(k_{n+m}h)\omega_{n+m+p}-g\frac{{\bf k}_n\cdot{\bf k}_p+{\bf k}_m\cdot{\bf k}_p}{\omega_{p}}\right) \notag\\
	&+\frac{1}{6}F_{n+p}\left( k_{n+p}\tanh(k_{n+p}h)\omega_{n+m+p}-g\frac{{\bf k}_n\cdot{\bf k}_m+{\bf k}_p\cdot{\bf k}_m}{\omega_{m}}\right) \notag\\
	&+\frac{1}{6}F_{m+p}\left( k_{m+p}\tanh(k_{m+p}h)\omega_{n+m+p}-g\frac{{\bf k}_m\cdot{\bf k}_n+{\bf k}_p\cdot{\bf k}_n}{\omega_{n}}\right)\\	 
	 \mathcal{C}_{n+m+p}=&\frac{g}{24}\left(\frac{k_n^2}{\omega_n}+\frac{k_m^2}{\omega_m}+\frac{k_p^2}{\omega_p} \right)\notag\\
	&+\frac{1}{6}\left( \omega_nE_{m+p}+\omega_mE_{n+p}+\omega_pE_{n+m}\right)\notag\\
	&+\frac{1}{6}\bigg( F_{n+m}k_{n+m}\tanh(k_{n+m}h)\notag\\
	&+F_{n+p}k_{n+p}\tanh(k_{n+p}h)+F_{m+p}k_{m+p}\tanh(k_{m+p}h)\bigg) 
	\end{align}
\end{subequations}
Similarly, the other third-order transfer coefficients can be determined by switching the arguments in (\ref{eq:ABCnmp}) and are functions of the following arguments:
\begin{subequations}\label{eq:thirdTranfun}
	\begin{align}
	\mathcal{A}_{n\pm m\pm p}=\varLambda_2 &\left\lbrace\left[ \omega_{n},{\bf k}_n,k_n\right];\left[  \pm\omega_{ m},\pm{\bf k}_{m},k_m\right];\left[  \pm\omega_{ p},\pm{\bf k}_{p},k_p\right];\right.  \notag\\
	&\; \left[k_{n\pm m},E_{n\pm m},F_{n\pm m}\right] ;\left[ k_{n\pm p},E_{n\pm p},F_{n\pm p}\right] ;\notag\\
	&\,\left.\left[ k_{\pm m\pm p},E_{\pm m\pm p},F_{\pm m\pm p}\right];\left[ \omega_{n\pm m\pm p}\right]  \right\rbrace      \\[3pt]
	\mathcal{B}_{n\pm m\pm p}=\varGamma_2&\left\lbrace\left[ \omega_{n},{\bf k}_n,k_n\right];\left[  \pm\omega_{ m},\pm{\bf k}_{m},k_m\right];\left[  \pm\omega_{ p},\pm{\bf k}_{p},k_p\right];\right. \notag\\
	& \;\left[k_{n\pm m},E_{n\pm m},F_{n\pm m}\right] ;\left[ k_{n\pm p},E_{n\pm p},F_{n\pm p}\right] ;\notag\\
	&\,\left.\left[ k_{\pm m\pm p},E_{\pm m\pm p},F_{\pm m\pm p}\right];\left[ \omega_{n\pm m\pm p}\right]  \right\rbrace \\[3pt]
	\mathcal{C}_{n\pm m\pm p}=\varXi_2&\left\lbrace\left[ \omega_{n},{\bf k}_n,k_n\right];\left[  \pm\omega_{ m},\pm{\bf k}_{m},k_m\right];\left[  \pm\omega_{ p},\pm{\bf k}_{p},k_p\right];\right. \notag\\
	& \;\left[k_{n\pm m},E_{n\pm m},F_{n\pm m}\right] ;\left[ k_{n\pm p},E_{n\pm p},F_{n\pm p}\right] ;\notag\\
	&\,\left.\left[ k_{\pm m\pm p},E_{\pm m\pm p},F_{\pm m\pm p}\right] \right\rbrace 
	\end{align}
\end{subequations}

\section{Derivation of coefficients in Hamiltonian solutions}\label{appB}
\subsection{Coefficients of canonical transformation}
From (\ref{eq:midevolutionA}) to the Zakharov equation (\ref{eq:Zakharov}). one can obtain the second-order coefficients
\footnotesize
\begin{equation}
  A_{0,1,2}^{(1)}=-\varDelta_{0-1-2}^{-1}U_{0,1,2}^{(1)},\quad A_{0,1,2}^{(2)}=-2\varDelta_{0+1-2}^{-1}U_{2,1,0}^{(1)}=-2A_{2,1,0}^{(1)},\quad A_{0,1,2}^{(3)}=-\varDelta_{0+1+2}^{-1}U_{0,1,2}^{(3)},
\end{equation}
\normalsize
and the third-order coefficients
\begin{subequations}
\begin{align}
  B_{0,1,2,3}^{(1)}&=-\varDelta_{0-1-2-3}^{-1}(Z^{(1)}_{0,1,2,3}+V^{(1)}_{0,1,2,3}),\\[3pt]
  T_{0,1,2,3}&=\varDelta_{0+1-2-3}B_{0,1,2,3}^{(2)}+Z^{(2)}_{0,1,2,3}+V^{(2)}_{0,1,2,3}\label{subeq:T0123a},\\[3pt]
  B_{0,1,2,3}^{(3)}&=-\varDelta_{0+1+2-3}^{-1}(Z^{(3)}_{0,1,2,3}+3V^{(1)}_{3,2,1,0}),\\[3pt]
  B_{0,1,2,3}^{(4)}&=-\varDelta_{0+1+2+3}^{-1}(Z^{(4)}_{0,1,2,3}+V^{(4)}_{0,1,2,3}),
\end{align}
\end{subequations}
where
  \begin{align*}
    Z_{0,1,2,3}^{(1)}&=\frac{2}{3}\left[U_{0,1,0-1}^{(1)}A_{2+3,2,3}^{(1)}+ U_{0,2,0-2}^{(1)}A_{1+3,1,3}^{(1)}+U_{0,3,0-3}^{(1)}A_{1+2,1,2}^{(1)}\right. \\
    &\quad\left.+U_{2,0,2-0}^{(1)}A_{-1-3,1,3}^{(3)}+ U_{3,0,3-0}^{(1)}A_{-1-2,1,2}^{(3)}+U_{1,0,1-0}^{(1)}A_{-2-3,2,3}^{(3)} \right] \\[3pt]
    Z_{0,1,2,3}^{(2)}&=-2\left[U_{0,2,0-2}^{(1)}A_{3,1,3-1}^{(1)}+ U_{0,3,0-3}^{(1)}A_{2,1,2-1}^{(1)}+U_{2,0,2-0}^{(1)}A_{1,3,1-3}^{(1)}\right. \\
    &\quad\left.+U_{3,0,3-0}^{(1)}A_{1,2,1-2}^{(1)}- U_{0+1,0,1}^{(1)}A_{2+3,2,3}^{(1)}-U_{-0-1,0,1}^{(3)}A_{-2-3,2,3}^{(3)} \right] \\[3pt]
    Z_{0,1,2,3}^{(3)}&=2\left[U_{0,3,0-3}^{(1)}A_{-1-2,1,2}^{(3)}- U_{0+1,0,1}^{(1)}A_{3,2,3-2}^{(1)}-U_{0+2,0,2}^{(1)}A_{3,1,3-1}^{(1)}\right. \\
    &\quad\left.+U_{3,0,3-0}^{(1)}A_{1+2,1,2}^{(1)}- U_{0,2,-0-2}^{(3)}A_{1,3,1-3}^{(1)}-U_{-0-1,0,1}^{(3)}A_{2,3,2-3}^{(1)} \right] \\[3pt]
    Z_{0,1,2,3}^{(4)}&=\frac{2}{3}\left[U_{-0-1,0,1}^{(3)}A_{2+3,2,3}^{(1)}+ U_{-0-2,0,2}^{(3)}A_{1+3,1,3}^{(1)}+U_{-0-3,0,3}^{(3)}A_{1+2,1,2}^{(1)}\right. \\
    &\quad\left.+U_{0+2,0,2}^{(1)}A_{-1-3,1,3}^{(3)}+ U_{0+3,0,3}^{(1)}A_{-1-2,1,2}^{(3)}+U_{0+1,0,1}^{(1)}A_{-2-3,2,3}^{(3)} \right]. 
 \end{align*}
The coefficient $B_{0,1,2,3}^{(2)}$ and the Zakharov kernel $T_{0,1,2,3}$ require special treatments due to the singularity caused by quartet resonance. Using the symmetry property of $T_{0,1,2,3}=T_{3,2,1,0}$, we obtain the canonicity condition for $B_{0,1,2,3}^{(2)}$:
\begin{equation}\label{eq:canonicityconditionT}
  \varDelta_{0+1-2-3}(B_{0,1,2,3}^{(2)}+B_{3,2,1,0}^{(2)})+Z_{0,1,2,3}^{(2)}-Z_{3,2,1,0}^{(2)}=0
\end{equation}
It means that $B_{0,1,2,3}^{(2)}$ and $T_{0,1,2,3}$ cannot be determined uniquely from \eqref{eq:canonicityconditionT}. A particular solution is given by
\begin{equation}\label{eq:B20}
  B_{0,1,2,3}^{(2)}=-\frac{1}{4}\varDelta_{0+1-2-3}^{-1}\left[ 3Z_{0,1,2,3}^{(2)}-Z_{1,0,2,3}^{(2)}-Z_{2,3,0,1}^{(2)}-Z_{3,2,0,1}^{(2)}\right] 
\end{equation}
Substituting $Z^{(2)}$ in the above equation gives 
\begin{align}
	B_{0,1,2,3}^{(2)}=&-\varDelta_{0+1-2-3}^{-1}\times\nonumber\\
	&U^{(1)}_{0,2,0-2}U^{(1)}_{3,1,3-1}\left(\frac{1}{\omega_3-\omega_1-\omega_{3-1}}- \frac{1}{\omega_0-\omega_2-\omega_{0-2}}\right)\nonumber\\
	&U^{(1)}_{0,3,0-3}U^{(1)}_{2,1,2-1}\left(\frac{1}{\omega_2-\omega_1-\omega_{2-1}}- \frac{1}{\omega_0-\omega_3-\omega_{0-3}}\right)\nonumber\\
	&U^{(1)}_{2,0,2-0}U^{(1)}_{1,3,1-3}\left(\frac{1}{\omega_1-\omega_3-\omega_{1-3}}- \frac{1}{\omega_2-\omega_0-\omega_{2-0}}\right)\nonumber\\
	&U^{(1)}_{3,0,3-0}U^{(1)}_{1,2,1-2}\left(\frac{1}{\omega_1-\omega_2-\omega_{1-2}}- \frac{1}{\omega_3-\omega_0-\omega_{3-0}}\right)\nonumber\\
	&U^{(1)}_{0+1,0,1}U^{(1)}_{2+3,2,3}\left(\frac{1}{\omega_{0+1}-\omega_0-\omega_1}-\frac{1}{\omega_{2+3}-\omega_2-\omega_3}\right)\nonumber\\
	&U^{(3)}_{-0-1,0,1}U^{(3)}_{-2-3,2,3}\left(\frac{1}{\omega_{0+1}+\omega_0+\omega_1}-\frac{1}{\omega_{2+3}+\omega_2+\omega_3}\right)
\end{align}
Fortunately, the terms in brackets from the above equation are all proportional to $\varDelta_{0+1-2-3}$. For example, the first term becomes
\begin{equation*}
	\frac{1}{\omega_3-\omega_1-\omega_{3-1}}- \frac{1}{\omega_0-\omega_2-\omega_{0-2}}=\frac{\varDelta_{0+1-2-3}+\omega_{3-1}-\omega_{0-2}}{(\omega_3-\omega_1-\omega_{3-1})(\omega_0-\omega_2-\omega_{0-2})}.
\end{equation*}
When approaching quartets resonance criterion, the term $\omega_{3-1}-\omega_{0-2}$ vanishes and $\varDelta_{0+1-2-3}$ is just cancelled. Finally, one obtain 
\begin{align}\label{eq:B2}
  B_{0,1,2,3}^{(2)}=&A_{0,1,-0-1}^{(3)}A_{2,3,-2-3}^{(3)}+A_{1,2,1-2}^{(1)}A_{3,0,3-0}^{(1)}+A_{1,3,1-3}^{(1)}A_{2,0,2-0}^{(1)}\nonumber\\
  &\quad-A_{0+1,0,1}^{(1)}A_{2+3,2,3}^{(1)}-A_{0,2,0-2}^{(1)}A_{3,1,3-1}^{(1)}-A_{0,3,0-3}^{(1)}A_{2,1,2-1}^{(1)}
\end{align}
The new expression of $B_{0,1,2,3}^{(2)}$ gets rid of the singularity $\varDelta_{0+1-2-3}=0$, and it means the complete separation between resonant modes and bound modes. Then, substituting \eqref{eq:B2} in \eqref{subeq:T0123a}, we obtain the Zakharov kernel
\begin{align}
  T_{0,1,2,3}&=\frac{1}{4}(Z_{0,1,2,3}^{(2)}+Z_{1,0,2,3}^{(2)}+Z_{2,3,0,1}^{(2)}+Z_{3,2,0,1}^{(2)})+V_{0,1,2,3}^{(2)}\notag\\
  &=V^{(2)}_{0,1,2,3}\notag\\
  &-U^{(1)}_{0,2,0-2}U^{(1)}_{3,1,3-1}\left(\frac{1}{\omega_2+\omega_{0-2}-\omega_0}+ \frac{1}{\omega_1+\omega_{3-1}-\omega_3}\right) \notag\\
  &-U^{(1)}_{1,3,1-3}U^{(1)}_{2,0,2-0}\left(\frac{1}{\omega_0+\omega_{2-0}-\omega_2}+ \frac{1}{\omega_3+\omega_{1-3}-\omega_1}\right) \notag\\
  &-U^{(1)}_{1,2,1-2}U^{(1)}_{3,0,3-0}\left(\frac{1}{\omega_2+\omega_{1-2}-\omega_1}+ \frac{1}{\omega_0+\omega_{3-0}-\omega_3}\right) \notag\\
  &-U^{(1)}_{0,3,0-3}U^{(1)}_{2,1,2-1}\left(\frac{1}{\omega_1+\omega_{2-1}-\omega_2}+ \frac{1}{\omega_3+\omega_{0-3}-\omega_0}\right) \notag\\
  &-U^{(1)}_{0+1,0,1}U^{(1)}_{2+3,2,3}\left(\frac{1}{\omega_{0+1}-\omega_{0}-\omega_1}+ \frac{1}{\omega_{2+3}-\omega_{2}-\omega_3}\right) \notag\\
  &-U^{(3)}_{-0-1,0,1}U^{(3)}_{-2-3,2,3}\left(\frac{1}{\omega_{0+1}+\omega_{0}+\omega_1}+ \frac{1}{\omega_{2+3}+\omega_{2}+\omega_3}\right) 
\end{align}

Futhermore, in \eqref{eq:phivspsi} the coefficients $C^{()}$ are given by 
\begin{subequations}
	\begin{align}
		C_{0,1,2,3}^{(1)}=&\frac{1}{3}\left( q_1M_1N_{0-1}A_{2+3,2,3}^{(1)}+q_1M_1N_{0-1}A_{-2-3,2,3}^{(3)}+q_{0-1}M_{0-1}N_1A_{2+3,2,3}^{(1)}\right.\nonumber\\
		&+q_2M_2N_{0-2}A_{1+3,1,3}^{(1)}+q_2M_2N_{0-2}A_{-1-3,1,3}^{(3)}+q_{0-2}M_{0-2}N_2A_{1+3,1,3}^{(1)} \nonumber\\
		&+q_3M_3N_{0-3}A_{1+2,1,2}^{(1)}+q_3M_3N_{0-3}A_{-1-2,1,2}^{(3)}+q_{0-3}M_{0-3}N_3A_{1+2,1,2}^{(1)} \nonumber\\
		&\left. +D_{0,1,2,3}^{(3)}M_1N_2N_3+D_{0,2,1,3}^{(3)}M_2N_1N_3+D_{0,3,2,1}^{(3)}M_3N_2N_1\right) \\
		C_{0,1,2,3}^{(2)}=&-q_3M_3N_{0-3}A_{2,1,2-1}^{(1)}-q_3M_3N_{0-3}A_{1,2,1-2}^{(1)}\nonumber\\
		&-q_2M_2N_{0-2}A_{3,1,3-1}^{(1)}-q_2M_2N_{0-2}A_{1,3,1-3}^{(1)}\nonumber\\
		&-q_{0-3}M_{0-3}N_3A_{2,1,2-1}^{(1)}-q_{0-2}M_{0-2}N_2A_{3,1,3-1}^{(1)}\nonumber\\
		&+q_{0+1}M_{0+1}N_1A_{2+3,2,3}^{(1)}+D_{0,3,2,-1}^{(3)}M_3N_2N_1+D_{0,2,3,-1}^{(3)}M_2N_3N_1\\
		C_{0,1,2,3}^{(3)}=&q_3M_3N_{0-3}A_{-1-2,1,2}^{(3)}+q_3M_3N_{0-3}A_{1+2,1,2}^{(1)}\nonumber\\
		&-q_{0+1}M_{0+1}N_1A_{3,2,3-2}^{(1)}-q_{0+2}M_{0+2}N_2A_{3,1,3-1}^{(1)}\nonumber\\
		&+q_{0-3}M_{0-3}N_3A_{-1-2,1,2}^{(3)}+D_{0,3,-2,-1}^{(3)}M_3N_2N_1 \\
		C_{0,1,2,3}^{(4)}=&\frac{1}{3}\left(q_{0+3}M_{0+3}N_3A_{-1-2,1,2}^{(3)}\right. \nonumber\\ 
		&\left. +q_{0+1}M_{0+1}N_1A_{-2-3,2,3}^{(3)}+ q_{0+2}M_{0+2}N_2A_{-1-3,1,3}^{(3)}\right) 		
	\end{align}
\end{subequations}
\subsection{Nonlinear transfer coefficients}
The second-order coefficients are
\begin{equation}
  U_{0,1,2}^{(1)}=-U_{-0,1,2}+U_{2,1,-0}-U_{-0,2,1},\quad U_{0,1,2}^{(3)}= U_{ 0,1,2}+U_{2,1,0}+U_{0,2,1},
\end{equation}
and the third-order coefficients are
\begin{subequations}
	\begin{align}
	V_{0,1,2,3}^{(1)}&=\frac{1}{3}(-V_{-0,1,2,3}+V_{2,1,-0,3}-V_{-0,2,1,3}+V_{3,1,2,-0}-V_{-0,3,2,1}+V_{2,3,1,-0}),\\
	V_{0,1,2,3}^{(2)}&=V_{-0,-1,2,3}-V_{2,-1,-0,3}-V_{-0,2,-1,3}-V_{3,-1,2,-0}-V_{-0,3,2,-1}+V_{2,3,-1,-0},\\
	V_{0,1,2,3}^{(4)}&=\frac{1}{3}(V_{0,1,2,3}+V_{2,1,0,3}+V_{0,2,1,3}+V_{3,1,2,0}+V_{0,3,2,1}+V_{2,3,1,0}),
	\end{align}
\end{subequations}
where
\begin{equation}
  U_{0,1,2}=-M_0M_1N_2E_{0,1,2}^{(3)},\quad V_{0,1,2,3}=-2M_0M_1N_2N_3E_{0,1,2,3}^{(4)}.
\end{equation}
Here, $E_{0,1,2}^{(3)}$ and $E_{0,1,2,3}^{(4)}$ have been given in \eqref{eq:Etranscoe}.

\bibliographystyle{jfm}
\bibliography{jfm-references}
\end{document}